\documentclass[fleqn,usenatbib]{mnras}
\usepackage{txfonts} 
\usepackage[T1]{fontenc}
\usepackage{graphicx}

\title{The Field White Dwarf Mass Distribution}

\author[Tremblay et al.]
{P.-E. Tremblay$^{1}$\thanks{E-mail: P-E.Tremblay@warwick.ac.uk},
J. Cummings$^{2}$,
J.~S. Kalirai$^{2,3}$,
B.~T. G\"ansicke$^{1}$,
\newauthor{N. Gentile-Fusillo$^{1}$, and R. Raddi$^{1}$}
\\
$^{1}$Department of Physics, University of Warwick, CV4 7AL, Coventry, UK\\
$^{2}$Center for Astrophysical Sciences, Johns Hopkins University, 3400
  North Charles Street, Baltimore, MD 21218, USA\\
$^{3}$Space Telescope Science
  Institute, 3700 San Martin Drive, Baltimore, MD 21218, USA
}
\volume{{\rm in press}}

\pubyear{2016}

\begin{document}
\label{firstpage}
\pagerange{\pageref{firstpage}--\pageref{lastpage}}
\maketitle

\begin{abstract}
  We revisit the properties and astrophysical implications of the field white
  dwarf mass distribution in preparation of \textit{Gaia} applications. Our
  study is based on the two samples with the best established completeness and
  most precise atmospheric parameters, the volume-complete survey within 20~pc
  and the Sloan Digital Sky Survey (SDSS) magnitude-limited sample. We explore
  the modelling of the observed mass distributions with Monte Carlo
  simulations, but find that it is difficult to constrain independently the
  initial mass function (IMF), the initial-to-final-mass relation (IFMR), the
  stellar formation history (SFH), the variation of the Galactic disk vertical
  scale height as a function of stellar age, and binary evolution. Each of
  these input ingredients has a moderate effect on the predicted mass
  distributions, and we must also take into account biases owing to
  unidentified faint objects (20~pc sample), as well as unknown masses for
  magnetic white dwarfs and spectroscopic calibration issues (SDSS
  sample). Nevertheless, we find that fixed standard assumptions for the above
  parameters result in predicted mean masses that are in good qualitative
  agreement with the observed values. It suggests that derived masses for both
  studied samples are consistent with our current knowledge of stellar and
  Galactic evolution. Our simulations overpredict by 40-50\% the number of
  massive white dwarfs ($M > 0.75$ $M_{\odot})$ for both surveys, although we
  can not exclude a Salpeter IMF when we account for all biases.  Furthermore,
  we find no evidence of a population of double white dwarf mergers in the
  observed mass distributions.

\end{abstract}


\section{Introduction}

Stars are an integral part of the luminous baryonic component of galaxies. As
a consequence, the IMF and SFH are important parameters in galactic evolution
models. Both quantities can be studied from a comparison of spectral
population synthesis models with samples of galaxies at different redshifts
\citep[see, e.g.,][]{maraston05,daddi07,conroy10}. Furthermore, resolved
stellar populations in the the Milky Way and nearby galaxies inform us more
directly on the IMF \citep[see, e.g.,][]{kroupa93,chabrier03,bastian10}. In
particular, recent Hubble Space Telescope observations of young clusters in
M31 \citep[see, e.g.,][]{imf} suggest that the high-mass IMF above $M \gtrsim
2 M_{\odot}$ is on average slightly steeper than the commonly used
\citet{salpeter} model.

Young and massive star clusters are ideal for studying the high-mass IMF since
they still include bright intermediate-mass stars ($1.5 \lesssim M/M_{\odot}
\lesssim 8$). However, the vast majority of these stars that were ever born in
the local galaxy group are now white dwarfs. While these faint remnants can
not be observed to the same distances as their progenitors, the field white
dwarf mass distribution could still provide information about the IMF of local
populations, such as the Galactic disk and halo. Current white dwarf samples
are small for Galactic halo studies but \textit{Gaia} \citep{carrasco14} and
\textit{Euclid} \citep{euclid} will soon present unprecedented
opportunities. Furthermore, the mass distribution of degenerate stars presents
unique constraints on the population of white dwarf mergers, which could be
one of the evolution channel linked to SN Ia \citep[see, e.g.,][]{sn1a} as
well as high-field magnetic white dwarfs \citep[B $>$ 1
MG;][]{tout,tout08,berro12,wickra}.  Extensive studies have also been dedicated to
using white dwarf masses and cooling ages to derive the SFH \citep{tremblay14}
and IFMR \citep[see,
e.g.,][]{IFMR1,catalan08,kalirai08,williams09,dobbie12,cummings15,cummings16}. The
IFMR provides constraints on the luminosity and lifetime of bright AGB stars,
hence on stellar population synthesis models \citep{marigo07,kalirai14}. While
these SFH and IFMR studies rely on individual stellar parameters, an
understanding of the overall field white dwarf mass distribution leads to an
essential internal consistency check for white dwarf and Galactic evolution
models.

Most of the $\sim$30,000 degenerate stars spectroscopically identified in the
SDSS \citep{alam15} have published masses
\citep{tremblay11,kleinman13,kepler16}, and among them $\sim$3000 have
sufficiently high signal-to-noise (S/N $\gtrsim 15$) to clearly define the
white dwarf mass distribution. In particular, the sample is large enough to
have outstanding statistics on less common high-mass white dwarfs (M $>$ 0.75
$M_{\odot}$). However, difficulties in the interpretation of the SDSS mass
distribution arise from significant spectroscopic selection biases as a
function of colour and magnitude \citep{gennaro08,fusillo15}. The local 20~pc
sample \citep{holberg08,sion09,gia12,sion14} offers a better completeness but
is limited to $\sim$120 objects.  In all cases, multiple astrophysical effects
have to be considered when attempting to comprehend field white dwarf masses,
such as the SFH, IMF, IFMR, and binary evolution.  Nevertheless, a few
attempts have been made at understanding features in the observed white dwarf
mass distributions \citep{ferrario05,catalan08,alberto11}. In the recent
years, most of these studies have been aimed at identifying a population of
mergers, though there are currently different interpretations on whether there
is evidence for merger products
\citep{liebert05,ferrario05,gia12,wegg12,isern13,alberto15}. To our knowledge,
there was no extensive study connecting white dwarf mass distributions to
Galactic archeology.  This is in contrast to white dwarf luminosity functions
that have been employed to derive the age and formation history of the
Galactic disk \citep[see, e.g.,][]{winget87,harris06,rowell13}.

The initial goal of this study was to constrain the IMF from white dwarf mass
distributions drawn from the 20~pc and SDSS samples.  We rapidly found out
that uncertainties in astrophysical relations and biases prevent a
straightforward interpretation, much in contrast with our earlier study of the
SFH from the local 20~pc white dwarf sample \citep{tremblay14}. In this work,
we present instead a systematic review of the uncertainties that come into
play when interpreting white dwarf mass distributions. Our investigation will
help to comprehend the larger \textit{Gaia} sample that will soon provide
precise individual luminosities for almost all known white dwarfs
\citep{carrasco14}. By combining parallaxes with spectroscopic or photometric
temperatures and a mass-radius relation, we will obtain precise masses
independently from current spectroscopic surface gravity measurements. We will
get much more precise mass distributions and gain a better understanding of
the completeness of current samples.

We base our study on Monte-Carlo simulations considering the IMF, IFMR, SFH,
main-sequence evolution, white dwarf evolution, galaxy kinematics, and survey
biases. We explore the modelling of both the observed 20~pc and SDSS mass
distributions. Since the white dwarfs studied in this work are restricted to
distances below $\sim$500 pc, we made no attempt at a full scale model of the
Galaxy. The assumptions behind our simulations are fairly similar to those
employed in recent studies of white dwarf luminosity functions and kinematics
\citep{wegg12,verbeek14,torres14,garcia2,lam15,alberto15}. As a consequence,
we concentrate on the often overlooked white dwarf mass distributions. In
Section~2 we describe our selected white dwarf samples. We continue with a
description of our standard simulations in Section~3 followed by a lengthy
discussion of uncertainties (Section~4). We comment on the results in
Section~5 and conclude in Section~6.

\section{White Dwarf Samples}

The absolute magnitude of a white dwarf is strongly dependent on its mass and
age. In order to study mass distributions, it is therefore essential to have a
clear assessment of the completeness of the observed samples. As a
consequence, we restrict our study to two white dwarf samples whose
completeness has been extensively characterised. Those correspond to the
volume-complete 20 pc survey, as well as the largely magnitude-limited SDSS
sample. Below $M_{\rm WD} \sim 0.45~M_{\sun}$, all objects are thought to be
He-core white dwarfs that are the product of close binary evolution. Assuming
single star evolution, these objects would have main-sequence lifetimes that
are longer than the age of the Universe. We found it practical to simply
remove those objects from the comparison of the observed and predicted mass
distributions. Extensive population studies including binary evolution have
already been performed \citep[see, e.g.,][]{binary2,binary1} and it is outside
the scope of our investigation to review these results. Throughout this work,
all our quoted values are for white dwarf masses above 0.45~$M_{\sun}$. The
main properties of our observed samples are identified in Table~\ref{fg:t0}
and described below.

\subsection{20~pc Sample}

We rely on the local 20 pc sample as presented in Table~\ref{fg:t2} of
\citet{gia12} with their corrections for 3D convective effects (see their
fig. 16). We have removed objects with estimated distances above 20~pc as well
as 12 members with no mass determination for a total of 105 white dwarfs,
among them 97 with $M>0.45M_{\odot}$. The atmospheric parameters were
determined from a combination of photometric, spectroscopic, and parallax
observations. In general, the combination of the photometric fluxes with
parallax allowed for the most precise luminosity and effective temperature
($T_{\rm eff}$) determinations. The masses were then derived employing the
evolutionary models of \citet{fontaine01}. These models have C/O cores (50/50
by mass fraction mixed uniformly) and assume thick hydrogen layers ($M_{\rm
  H}/M_{\rm total}$ = 10$^{-4}$) for H-atmosphere white dwarfs and thin layers
($M_{\rm H}/M_{\rm total}$ = 10$^{-10}$) for helium and mixed atmospheres.

\begin{table}
%\scriptsize
\centering
\caption{Observed Samples}
\setlength{\tabcolsep}{3.25pt}
\begin{tabular}{@{}lll}
\hline
\hline
\noalign{\smallskip}
 \multicolumn{1}{c}{Property} & 20~pc sample & SDSS sample\\

 \noalign{\smallskip}
 \hline
 \noalign{\smallskip}
Sample definition & \citealt{gia12} & \citealt{kleinman13}\\
Data reduction & Various & SDSS DR10\\
$T_{\rm eff}$/$\log g$ & \citealt{gia12} & This work\\
Cooling models & \citealt{fontaine01} & \citealt{fontaine01} \\
$T_{\rm eff}$ range (K) & Unrestricted & $16,000 < T_{\rm eff} < 22,000$\\
Magnitude range & Unrestricted & $16.0 < g < 18.5$\\
Mass range ($M_{\odot}$) & $>$ 0.45 & $>$ 0.45\\
Distance range (pc) & $<$ 20 & Unrestricted\\
Spectral types & Unrestricted & DA(Z) and DB(A)(Z)\\
Number & 97 & 715\\
Number (non-DA) & 34 & 135\\
Number (magnetic) & 15 & 0\\

 \noalign{\smallskip}
 \hline
 \noalign{\smallskip}
 \label{fg:t0}
\end{tabular}
\end{table}

\begin{figure}
  \centering \includegraphics[scale=0.475,bb=30 147 572 589]{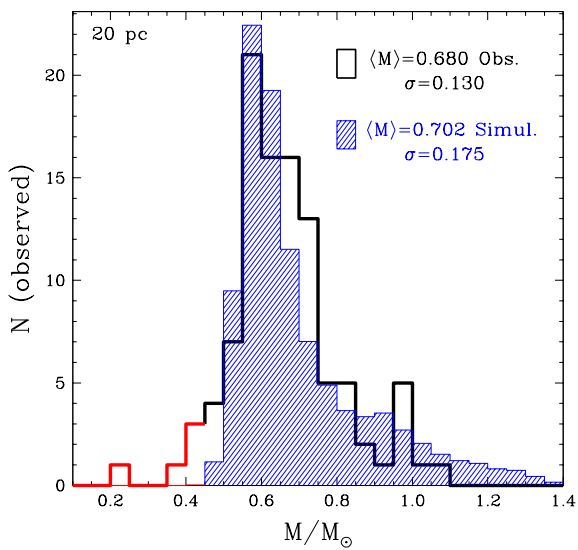}
  \caption[f1.eps]{Observed (black) and simulated (filled blue) mass
    distributions for the 20~pc sample of \citet{gia12}. The standard Monte
    Carlo simulation $A_{\rm 20pc}$ is described in Section~3. We neglect
    low-mass objects (red; $M < 0.45~M_{\odot}$) for the computation of the
    mean masses and mass dispersions, which are labeled on the
    panel. This should taken into account when comparing to other 
		studies that do not restrict mass values. \label{fg:f1}}
\end{figure}

We present the observed mass distribution in Fig.~\ref{fg:f1}. We remind
the reader that by restricting the selected stars to $M > 0.45~M_{\odot}$
the mean mass value is biased towards a higher value than those reported in other studies.
This sample is estimated to be 80-90\% volume-complete \citep{holberg08,gia12,tremblay14}.
At zeroth order, the mass of a white dwarf is expected to have little
dependence on the volume in which it is located. However, we discuss in
Section~4 that biases owing to variable SFH and velocity dispersion (as a
function of age and mass) have a significant impact on the interpretation of
the observed mass distribution. We also pay special attention to the fact that
the 10-20\% missing objects could preferentially be lower luminosity, hence
massive white dwarfs. We note that owing to the photometric technique which
has little sensitivity to the atmospheric composition, this sample has precise
masses for all spectral types, including DA, DB, DC, DQ, and DZ white dwarfs.
This also includes magnetic objects accounting for 15\% of the sample, 
which should be regarded as a lower limit given that many local white dwarfs
have not been adequately observed for polarisation and many of them are too 
cool to show Zeeman splitting.

\subsection{SDSS Sample}

We also rely on the SDSS white dwarf sample, which is largely
magnitude-limited but has a complex spectroscopic completeness that varies
from 10 to 90\% as a function of magnitude and colour
\citep{gennaro08,fusillo15}.  The other major difference with the 20~pc sample
is that no parallaxes are available for the vast majority of SDSS white
dwarfs. For DA and DB white dwarfs, which represent roughly 85\% of the SDSS
sample \citep{kleinman13}, it is possible to employ the spectroscopic method
combined with evolutionary sequences to determine the masses to a high
precision \citep{bergeron92,bergeron11}. On the other hand, it is not
straightforward to determine masses for other spectral types, in particular
magnetic DA white dwarfs. The resulting bias on the mass distribution could be
important, since magnetic degenerates are thought to be more massive than the
average \citep{ferrario15}.

We base our analysis on the SDSS Data Release 7 (DR7) spectroscopic sample of
\citet{kleinman13}, where we have carefully refitted all DAs with 1D
ML2/$\alpha$ = 0.8 model atmospheres \citep{tremblay11}, applied 3D
corrections \citep{tremblay13}, and re-assigned the different subtypes based
on a careful visual identification. We have employed SDSS spectra with the
data reduction from DR10. As for the 20~pc sample, we have employed
evolutionary models from \citet{fontaine01} to constrain
masses. \citet{tremblay11} have demonstrated that a cutoff at S/N $\sim$ 15
was an optimal separation between the size of the sample and the uncertainties
in the mass distribution, and we use a corresponding cutoff at $g < 18.5$.

\begin{figure}
  \centering \includegraphics[scale=0.475,bb=30 147 572 589]{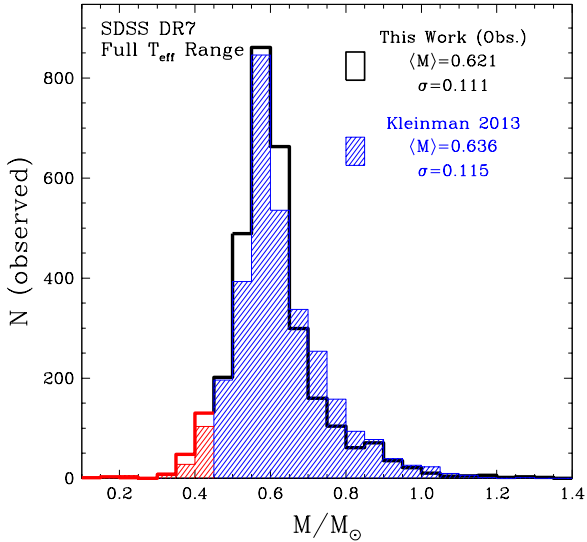}
  \caption[f2.eps]{Comparison of the SDSS DR7 mass distributions from this
    work (black) and \citet[][filled blue]{kleinman13} for single,
    non-magnetic DA white dwarfs with $g < 18.5$. The identification 
		of subtypes differs between the two studies, hence we renormalise
		the \citet{kleinman13} distribution of 2903 single non-magnetic DAs to the 2998 
		such objects identified in this work. We neglect low-mass objects
     (red; $M < 0.45~M_{\odot}$) for the computation of the mean masses and mass
    dispersions. Our reported mean masses are thus biased towards higher values
		than those previously published. \label{fg:f2}}
\end{figure}

The full results of our alternative analysis are outside the scope of this
work, and in Fig.~\ref{fg:f2} we simply compare our mass distribution with
that of \citet{kleinman13}. We restrict the comparison to single non-magnetic
DAs ($M < 0.45~M_{\odot}$) and we note that the identification of subtypes differs between the two
studies. We find a moderate offset of 0.015 $M_{\odot}$ in the mean
mass. Furthermore, our identification of subtypes is significantly different,
especially for magnetic white dwarfs and DA+DC double degenerates, which can
result in spurious high $\log g$ values if incorrectly identified. For
instance, we find a magnetic fraction of 2.5\% compared to 4.4\%-5.3\% for
\citet{kleinman13}, with their upper estimate based on uncertain
detections. Our recovered fraction is admittedly small, 6 times less than for
the 20~pc sample discussed above, and we have no explanation for this
behaviour. The true magnetic incidence in the SDSS is expected to be slightly
larger given the low average signal-to-noise of the spectroscopic
observations, but this can not account for the full difference. We are aware
that larger and more recent SDSS samples have since been identified, though
until the discrepancy with the identification of subtypes has been resolved,
we prefer to rely on our own sample. Furthermore, our study below is limited
by biases rather than number statistics, hence it was deemed unnecessary to
build a larger sample.

We complement our DA sample with the DB white dwarfs identified by
\citet{kleinman13}, using the atmospheric parameters recently determined by
\citet{koester15}. For this study, we rely only on the single DA and DB white
dwarfs (including objects with trace elements such as DAZ or DBA), hence we
remove stellar remnants with main-sequence companions, double degenerates, and
magnetic white dwarfs.  We nevertheless consider the effect of these missing
subtypes in our review of uncertainties in Section~4.

\begin{figure}
  \centering \includegraphics[scale=0.440,bb=20 25 592 730]{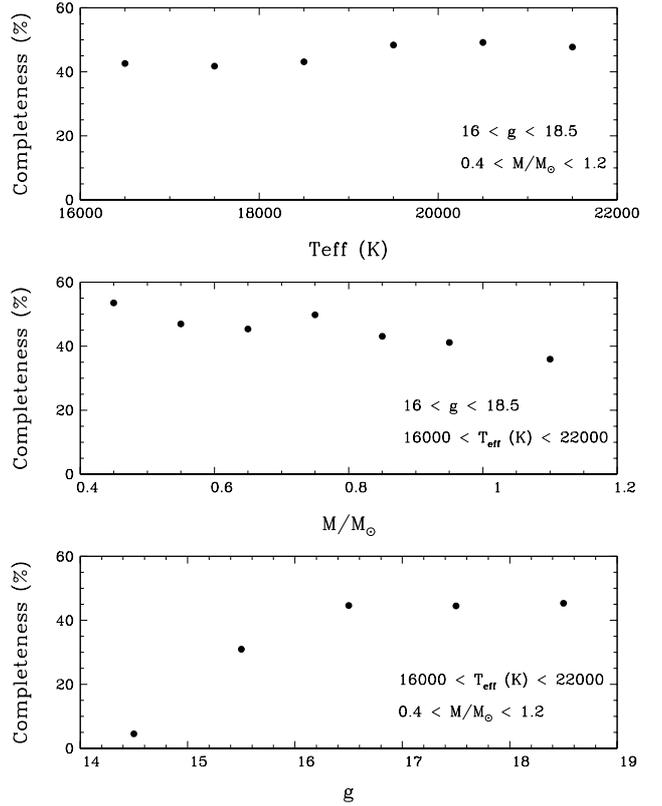}
  \caption[f3.eps]{Spectroscopic completeness of the SDSS DR10 sample as a function of
    $T_{\rm eff}$ (top panel), $M/M_{\odot}$ (middle), and $g$ magnitude
    (bottom) for DA white dwarfs. We only cover the $T_{\rm eff}$ range
    studied in this work. For objects that have a probability of 41\% or
    higher of being a white dwarf in \citet{fusillo15}, we define the
    completeness as the fraction of objects with a SDSS spectrum. We have
    transformed the observed colours into atmospheric parameters with model
    atmospheres from \citet{tremblay11}.  \label{fg:f3}}
\end{figure}

Below $T_{\rm eff} \lesssim 12,000$~K, helium-rich DC, DQ, and DZ white dwarfs
become present in significant numbers, and for $T_{\rm eff} < 16,000$ K, DBs
have uncertain spectroscopic parameters \citep{bergeron11,koester15}. As a
consequence, we restrict our study to higher temperatures where the large
majority of SDSS white dwarfs have precise atmospheric parameters and are
either of the DA, DB, or DBA spectral type. Furthermore, the spectroscopic
completeness of the SDSS catalog varies significantly as a function of colour
and magnitude. To circumvent this problem, we have decided to restrict our
analysis of the mass distribution to objects with $16,000 < T_{\rm eff}$ (K)
$< 22,000$ and $16.0 < g < 18.5$. This corresponds to cooling ages in the
range $0.02 \lesssim \tau$ (Gyr) $\lesssim 1$. We demonstrate in
Fig.~\ref{fg:f3} that the completeness can be approximated as constant within
that range.  We base our calculations on the results of \citet{fusillo15} who
have determined the probability of SDSS DR10 photometric sources of being
white dwarfs based on colours and proper-motion. We have converted the
observed colours into atmospheric parameters for DA white dwarfs, and we
define the completeness as the fraction of objects with a spectrum among those
that have a probability of 41\% or larger of being a white dwarf. We note that
the spectroscopic completeness appears to decrease slightly for higher masses,
though one should be cautious about this result since we do not account for
possible photometric calibration offsets, reddening, and a confirmation that
objects without a spectrum are actually stellar remnants. We remind the reader
that Fig.~\ref{fg:f3} refers to the spectroscopic completeness for a
magnitude-limited sample, and not the volume completeness, which will be taken
into account in our simulations in Section 3. DB white dwarfs span a similarly
located but much smaller colour space in that $T_{\rm eff}$ range, hence we
assume that the completeness is the same as for DAs. We concur with
\citet{gennaro08} that spectroscopic completeness corrections do not
necessarily have a large effect on the relative mass distribution.

\begin{figure}
  \centering \includegraphics[scale=0.475,bb=30 147 572 589]{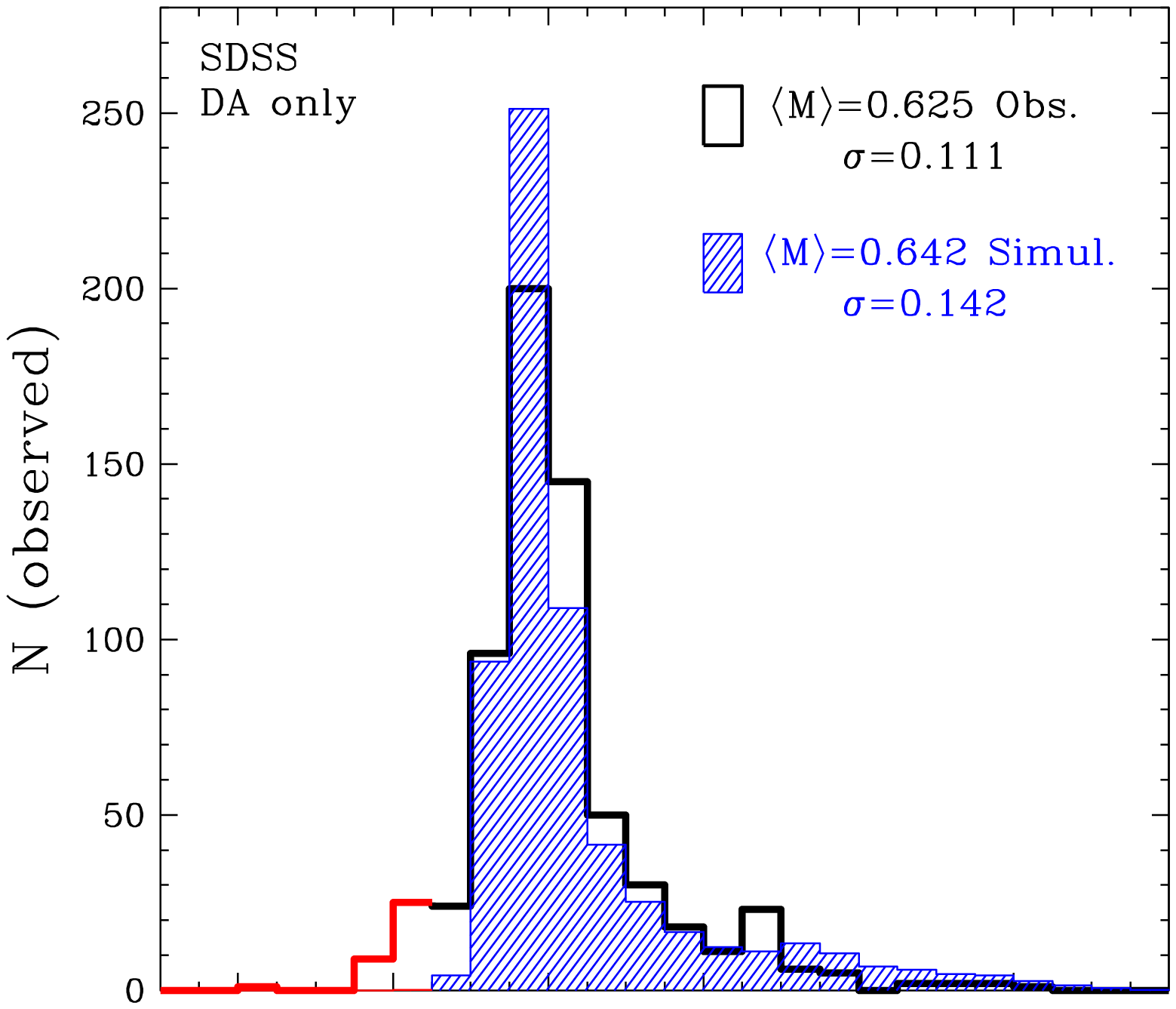}
  \centering \includegraphics[scale=0.475,bb=30 147 572 539]{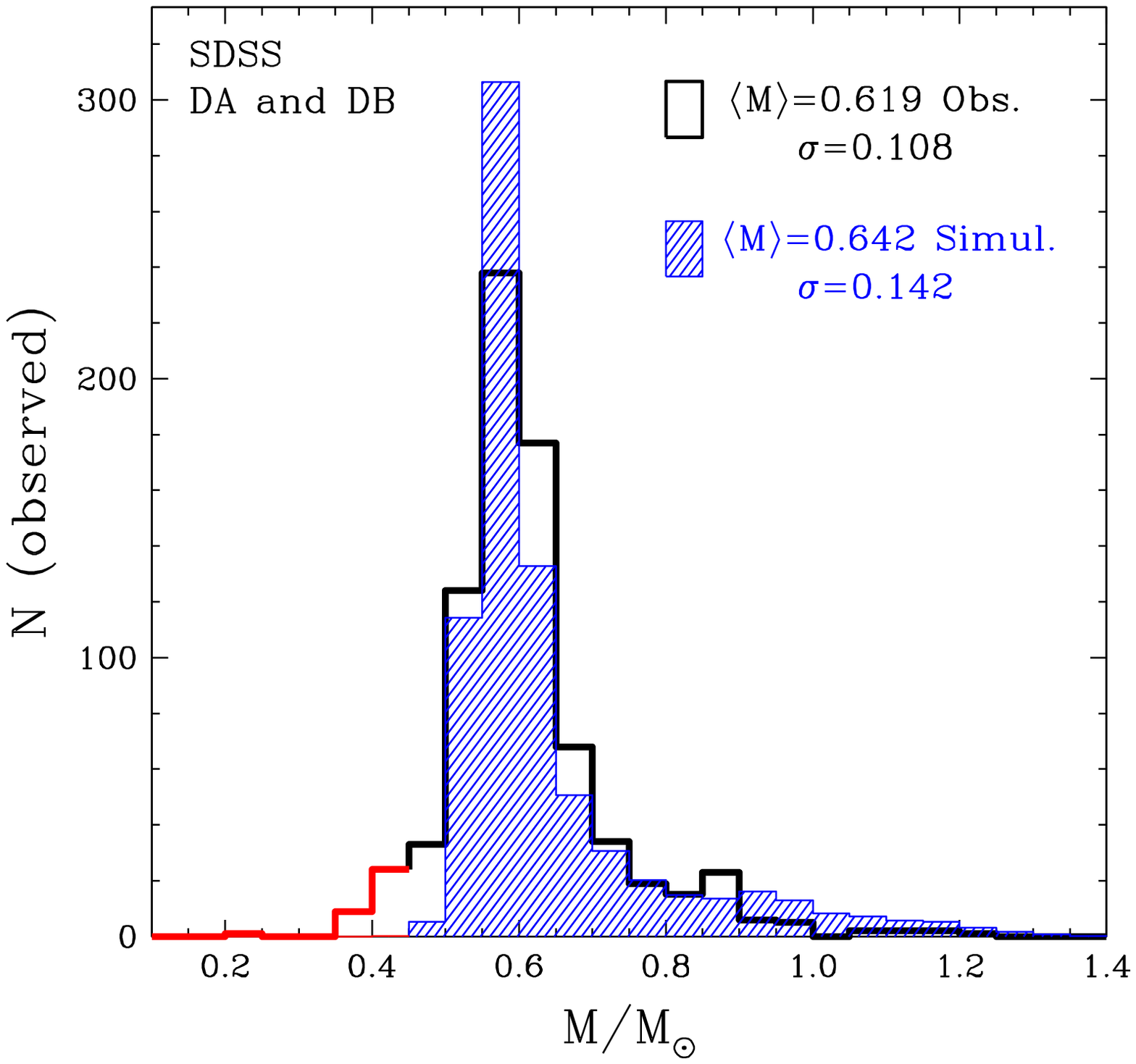}
  \caption[f4.eps]{{\it Top:} observed (black) and simulated (filled blue)
    mass distributions for the SDSS DR7 DA white dwarf sample for $16,000 <
    T_{\rm eff}$ (K) $< 22,000$ and $16.0 < g < 18.5$. The standard Monte
    Carlo simulation $A_{\rm SDSS}$ is described in Section~3. We have removed
    binaries and magnetic white dwarfs from the observed sample. We neglect
    low-mass objects (red; $M < 0.45~M_{\odot}$) for the computation of the
    mean masses and mass dispersions, which are labeled on the panel. {\it
      Bottom:} same as the top panel but for the combined DA and DB sample. \label{fg:f4}}
\end{figure}

Fig.~\ref{fg:f4} presents the mass distribution for our selected $T_{\rm eff}$
and $g$ magnitude range, both for DAs only and the combined DA and DB sample.
In the combined sample, the fraction of DBs is 18\%. We find that the DA mass
distribution (top panel) for our $T_{\rm eff}$ subsample is fairly similar to
the mass distribution for the full temperature range in Fig.~\ref{fg:f2}. We
study further our $T_{\rm eff}$ cutoffs in Section~4 and Table~\ref{fg:t2}. We
also observe that the addition of DB white dwarfs (bottom panel) leads to a
slightly smaller mean mass and mass dispersion. One reason is that there are
very few genuine high-mass helium-rich degenerate stars when uncertainties in
the line broadening are accounted \citep{bergeron11}. For the remainder of
this work, we employ the combined DA and DB sample.

\section{Simulations}

We have designed Monte Carlo simulations of white dwarf populations in the
solar neighborhood. The basic assumptions about these simulations are similar
to the ones presented in \citet{wegg12,garcia2,alberto15} and
\citet{torres16}. In this section, we present our standard model for the 20~pc
($A_{\rm 20 pc}$) and SDSS ($A_{\rm SDSS}$) samples, and defer the lengthy
discussion about biases and alternative assumptions to Section~4.

We use a simple galactic model with a thin disk that was formed 10 Gyr ago, a
constant star formation history (SFH), and a Salpeter initial mass function
($NdM_{\rm i}\propto M_{\rm i}^{-\alpha}dM_{\rm i}$, where $\alpha =
2.35$). We assume a uniform distribution in Galactic coordinates $U$ and $V$,
corresponding to the plane of the disk. This crude galactic model is a
reasonable assumption given that all of our targets are within a distance of
500~pc. We suppose that radial migration within the disk has no net effect on
the white dwarf mass distribution, which is also a consequence of assuming a
uniform distribution in the plane of the disk.

We employ a variable scale height in the vertical direction $W$ as a function
of total stellar age, the sum of the white dwarf cooling age and the
main-sequence lifetime.  We suppose that the vertical scale height is directly
proportional to the observed vertical velocity dispersion $\sigma_{\rm W}$, as
in the isothermal sheet galactic disk model \citep{spitzer42}. We use the
empirical velocity dispersion versus total age relation from \citet{vel}

\begin{equation}
\sigma_{\rm W} = k~{\rm age}^{0.6} ~~~~~ {\rm for~total~stellar~age} < 5 {\rm ~Gyr},
\end{equation}
\begin{equation}
\sigma_{\rm W} = k~{\rm 5}^{0.6} ~~~~~ {\rm for~total~stellar~age} > 5 {\rm ~Gyr},
\end{equation}

{\noindent}where $k$ is a constant and the total stellar age in Gyr. We have
chosen the constant so that the vertical scale height is 75~pc at 1~Gyr based
on the velocity of young massive SDSS white dwarfs \citep{wegg12} and the
scale height distribution of young open clusters \citep{buckner14}. The
vertical scale height thus reaches a maximum value of $\sim$200~pc at 5 Gyr
and thereafter remains constant according to Eqs.~1-2. Since white dwarfs have a
quite limited mass range, we do not consider a variation with mass. Furthermore,
our samples are centered 20~pc above the plane of the disk, which is the
approximate position of the Sun. 

The first step of our Monte Carlo simulations is to have a star formed at a
random time in the last 10~Gyr, a random mass weighted by the Salpeter IMF,
and a random location within 500~pc weighted by the exponential distribution
in the vertical direction. We then derive the main-sequence lifetime for solar
metallicity from \citet{hurley00}, and subtract it to the formation time to
obtain the white dwarf cooling age. If the cooling age is positive, we then
obtain the white dwarf cooling age. If the cooling age is positive, we then
find the white dwarf mass using an IFMR drawn from \citet{cummings16}
supplemented by low-mass data ($M_{\rm WD} < 0.65 M_{\odot}$) adapted from
\citet{kalirai07,kalirai08,kalirai09}.

\begin{equation}
M_{\rm WD}=0.541932-0.0184299 M_{\rm i}+0.0265180 M_{\rm i}^2 \\{\rm for~} M_{\rm i} < 4 M_{\odot}~,
\end{equation}

\begin{equation}
M_{\rm WD}=0.915738-0.0878731 M_{\rm i}+0.0208688 M_{\rm i}^2 \\{\rm for~} M_{\rm i} > 4 M_{\odot}~,
\end{equation}

{\noindent}where $M_{\rm WD}$ is the white dwarf mass and $M_{\rm i}$ is the
initial mass. We are aware that the observed IFMR is often represented as a
linear relation and that our two-part 2nd order fit may provide corrections to
the linear relation that are not physical. However, the IFMR is an empirical
relation and our 2nd order fit allows us to directly connect features in the
observed IFMR to features in the mass distribution \citep{ferrario05}. Unlike
the IMF, there is no theoretical suggestion that the IFMR should be a simple
analytical function. In fact, several theoretical IFMRs have a turnover at
$M_{\rm i} \sim 4 M_{\odot}$ resulting from the second dredge-up which only
occurs for higher masses \citep[see, e.g.,][]{marigo07,meng08}.

From the white dwarf mass and cooling age, we obtain $T_{\rm eff}$ and $\log
g$ with the evolution models of \citet{fontaine01}, as well as $V$ and $g$
magnitudes from the model atmospheres of \citet{tremblay11} and
\citet{bergeron11} for DA and DB white dwarfs, respectively.  We use the
observed number of H- and He-atmospheres. In all cases, we assume 70\% of
thick H-layers and 30\% of thin H-layers, which is based on the fact that a
fraction of DAs also have thin layers \citep{tremblay08}.

All further steps depend on the specific survey. In all cases, we let the
simulations run long enough so that 30,000 white dwarfs satisfying all cuts
are selected. We are only interested in relative numbers and we renormalise
our mass distributions to the actual number of observed white dwarfs with M >
0.45 $M_{\odot}$, unless simulations are compared.

\subsection{Simulations of the 20~pc Sample}

We have simulated observational errors with a Gaussian error distribution and
a 1$\sigma$ value based on the mean of the uncertainties given in
\citet{gia12}. It corresponds to 0.0375 $M_{\odot}$ in mass, 2.0\% in
temperature, and 0.7~pc in distance. The final selection is then made from all
objects within 20~pc. Fig.~\ref{fg:f1} compares our simulated mass
distribution with the observed one. We remind the reader that it is not a fit
and we made no attempt to tweak the input parameters of the simulation to
match the observations. We find that the agreement is quite good, both in
terms of the mean mass and the overall shape of the distribution. However, the
mass dispersion value and the number of massive white dwarfs are overpredicted
in the simulations.

\subsection{Simulations of the SDSS Sample}

\begin{figure}
  \centering \includegraphics[scale=0.475,bb=40 240 592 580]{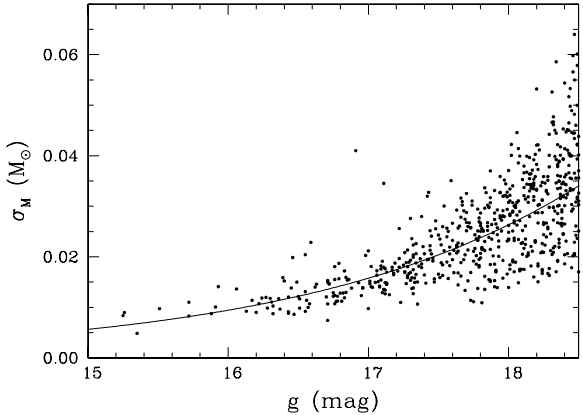}
  \caption[f5.eps]{Errors on derived masses for the
    SDSS DR7 DA white dwarf sample with $16,000 < T_{\rm eff}$ (K) $< 22,000$
    and $16.0 < g < 18.5$. The solid line is a fit to the observed
    distribution and represents the $1\sigma$ value of the Gaussian errors
    that we apply to our simulations.
    \label{fg:f5}}
\end{figure}

Fig.~\ref{fg:f5} illustrates that the mass error for DA white dwarfs in the
SDSS is strongly correlated with magnitude. We made a linear fit to the $\log
\sigma_{M}$ distribution, as shown in Fig.~\ref{fg:f5}, to determine the
1$\sigma$ value of our simulated Gaussian error distribution. We use similar
relations for temperature and magnitude errors, though those only have minor
roles compared to the mass errors. As discussed in Section~2.2, we then use a
cut given by $16,000 < T_{\rm eff}$ (K) $< 22,000$ and $16.0 < g < 18.5$, and
assume that within that range the completeness does not vary. Finally, the
SDSS DR7 sample is not isotropic in Galactic coordinates but mostly covers
high Galactic latitudes\footnote{http://classic.sdss.org/dr7/coverage/}. As a
consequence, we have employed coverage maps to select simulated objects that
are within the SDSS sky coverage. We note that the effect is similar to
changing the vertical scale height of the Galactic disk in our model.

Fig.~\ref{fg:f4} presents a comparison of the simulated and observed SDSS mass
distributions. The simulations for the DA only and combined DA and DB samples
are almost the same for this $T_{\rm eff}$ range.  It is shown that the
agreement between the observed and simulated distributions is similar to that
seen in Fig.~\ref{fg:f1} for the 20~pc sample. Once again, the number of
high-mass white dwarfs and the mass dispersion are predicted too large.

\subsection{Sample Comparison}

\begin{figure}
  \centering \includegraphics[scale=0.475,bb=30 147 572 589]{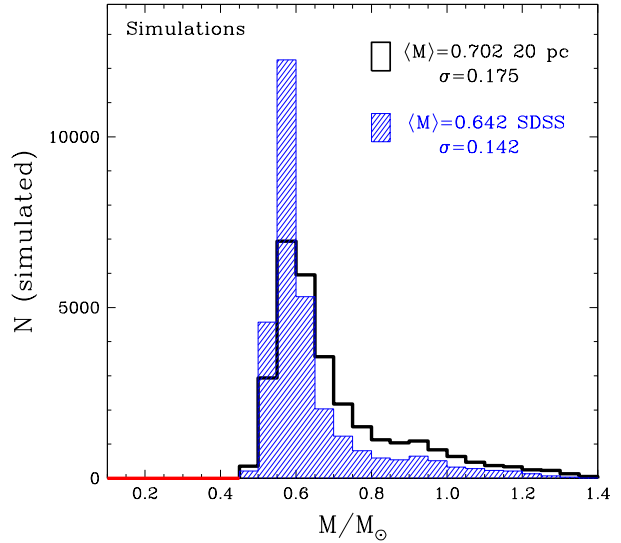}
  \caption[f8.eps]{Comparison of our standard Monte Carlo simulations $A_{\rm
      20pc}$ (black) and $A_{\rm SDSS}$ (filled blue) as previously shown in
    Figs.~\ref{fg:f1} and \ref{fg:f4}. There are 30,000 simulated objects in
    both samples to reduce statistical noise. \label{fg:f8}}
\end{figure}

The 20~pc sample is volume limited but the SDSS is largely magnitude limited,
hence their white dwarf mass distributions differ. In particular, one could
expect a smaller number of low luminosity massive white dwarfs in the SDSS
sample. Fig.~\ref{fg:f8} demonstrates that it is indeed the case, though
differences between both samples are more complex than just a simple
correction for stellar radius bias. We note that while the SDSS sample in
Fig.~\ref{fg:f8} is restricted to $16,000 < T_{\rm eff}$ (K) $< 22,000$ and
$16.0 < g < 18.5$, the mean mass differs by less than 1\% with no $T_{\rm
  eff}$ restriction.

\begin{figure}
  \centering \includegraphics[scale=0.425,bb=30 150 592 730]{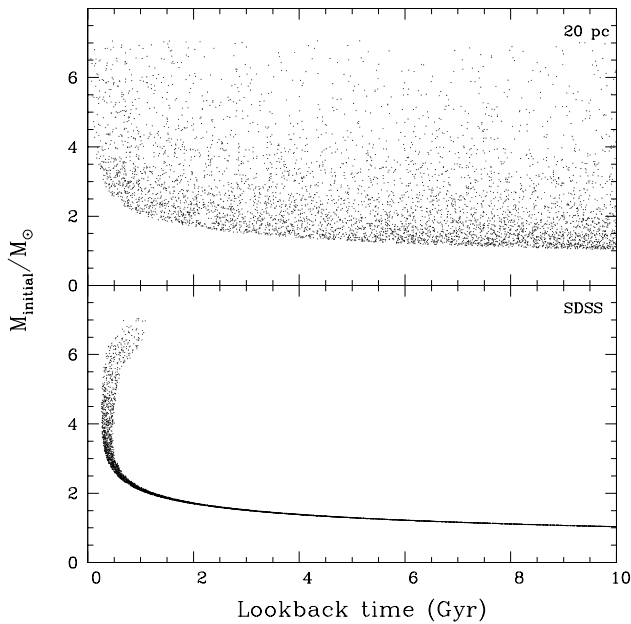}
  \caption[f6.eps]{Initial mass as a function of total stellar age for white
    dwarfs simulated in the 20~pc (top) and SDSS samples (bottom, $16,000 <
    T_{\rm eff}$ (K) $< 22,000$ and $16.0 < g < 18.5$).  There are 5000
    simulated objects in both samples.
    \label{fg:f6}}
\end{figure}

Fig.~\ref{fg:f6} shows the initial mass versus total age distribution for
white dwarfs in our simulated samples. The 20~pc sample covers a large range
of initial parameters, but our SDSS sample has a dramatically different
structure. The latter is limited to short-lived intermediate-mass stars ($2
\lesssim M/M_{\odot} \lesssim 8$) formed in the last 1 Gyr and older stars ($1
\lesssim M/M_{\odot} \lesssim 2$) that have just the right mass and age to
have recently evolved from the main-sequence 20-200 Myr ago. The steepness of
the main-sequence lifetime versus initial mass relation combined with white
dwarf cooling ages lead to a very specific coverage of initial parameters in
Fig.~\ref{fg:f6}. When this is joined with the IMF, IMFR, and variation of
vertical scale height as a function of stellar age, we can ultimately define
the number of high mass white dwarfs.  As a consequence, it is far from as
simple as correcting for stellar radius bias.

\begin{figure}
  \centering \includegraphics[scale=0.425,bb=30 390 592 730]{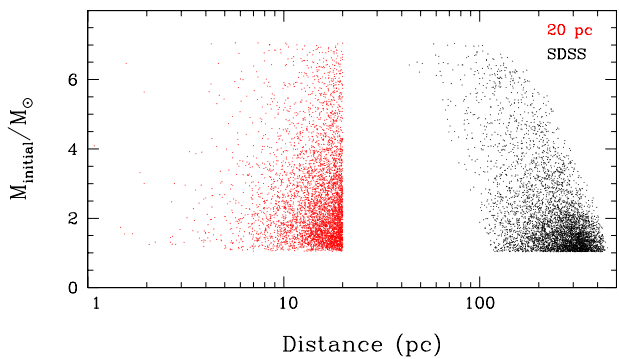}
  \caption[f7.eps]{Initial mass as a function of the distance to the Sun (logarithmic scale)
	  for white dwarfs simulated in the 20~pc (red) and SDSS samples (black, $16,000 <
    T_{\rm eff}$ (K) $< 22,000$ and $16.0 < g < 18.5$).
		There are 5000 simulated objects in both samples.
                \label{fg:f7}}
\end{figure}

The volume covered by our samples is shown in Fig.~\ref{fg:f7}. Volume effects
depend critically on the vertical scale height of the Galactic disk, which
varies from 75~pc at 1~Gyr to 200~pc at 5~Gyr in our model. For both samples,
this effect increases the number of high-mass white dwarfs in comparison to a
constant scale height. Indeed, massive white dwarfs come from short-lived
intermediate mass stars and have smaller total ages, hence a smaller average
vertical scale height. For the 20~pc sample which is located close to the
plane of the disk, it directly increases the number of high-mass white
dwarfs. For our SDSS sample, massive white dwarfs have total ages of $\sim$1
Gyr according to Fig.~\ref{fg:f6}, hence a vertical scale height of
$\sim$75~pc. Fig.~\ref{fg:f7} demonstrates that our sample is sensitive to
massive white dwarfs at distances of $\sim$50-150~pc, hence we maximize their
numbers by having a vertical scale height in the same range.

One important finding of this work is that the mean masses of the 20~pc and
SDSS white dwarf samples are predicted to be significantly different. We
predict a 9\% larger mean mass for the 20~pc sample, while observations in
Fig.~\ref{fg:f1} and Fig.~\ref{fg:f4} show an offset of 10\%. We therefore
suggest that this observed difference in mean mass is largely caused by the structure
of the samples, and not due to a systematic difference between the photometric
and spectroscopic techniques, which are the dominant methods to determine the
atmospheric parameters for the 20~pc and SDSS samples, respectively. We defer
to Section~5 the detailed comparison of the observed mass distributions and
simulations.

\section{Review of Uncertainties}

We now review in turn various uncertainties and biases that impact the
simulated mass distributions. As a summary of these experiments, we present in
Table~\ref{fg:t1} the mean mass, mass dispersion, and fraction of high-mass
white dwarfs for the observations and all numerical experiments (which we
label from $A$ to $P$). The fraction of high-mass white dwarfs is defined as
the fraction of objects with $M/M_{\odot} > 0.75$ with respect to the full
considered range ($M/M_{\odot} > 0.45$). For both surveys, the observed
fraction of massive remnants is too small compared to the predictions of our
standard simulations. This is in line with the preliminary assessment of
\citet{tremblay14} who found that the 20~pc mass distribution appeared
significantly steeper than the Salpeter IMF. For the 20~pc and SDSS samples,
the observed number of massive white dwarfs would need to increase by a factor
of 1.42 and 1.54, respectively, to match the simulations.

\begin{table*}
%\scriptsize
\centering 
\caption{Observed Mass Distributions and Monte Carlo Simulations}
\setlength{\tabcolsep}{3.25pt}
\begin{tabular}{@{}rcccccccc}
\hline
\hline
\noalign{\smallskip}
 & & \multicolumn{3}{c}{20~pc sample} & \multicolumn{3}{c}{SDSS sample} & \\
ID & \multicolumn{1}{c}{Data} & $\langle M \rangle$ & $\sigma_{\rm M}$ & N$(>0.75 M_{\odot})$ & $\langle M \rangle$ & $\sigma_{\rm M}$ & N$(>0.75 M_{\odot})$& Reference\\
 & \multicolumn{1}{c}{} & ($M_{\odot})$ & ($M_{\odot})$ & \% & ($M_{\odot})$ & ($M_{\odot})$ & \% &\\

 \noalign{\smallskip}
 \hline
 \noalign{\smallskip}
 $...$ & Observed                                 & 0.680 & 0.130 & 20.6 & 0.619 & 0.108 & 10.0 & Sect. 2, Figs.~\ref{fg:f1} and~\ref{fg:f4}\\
 \noalign{\smallskip}
 \hline
 \noalign{\smallskip}
 A  & Standard Monte Carlo                        & 0.702 & 0.175 & 26.9 & 0.642 & 0.142 & 14.6 & Sect. 3, Figs.~\ref{fg:f1} and~\ref{fg:f4}\\
 \noalign{\smallskip}
 \hline
 \noalign{\smallskip}
 B  & $V<17$ (20~pc only)                         & 0.687 & 0.162 & 24.3 &  $-$  &  $-$  &  $-$ & Sect. 4.1, Fig.~\ref{fg:f10}\\
 C  & No observational errors                     & 0.701 & 0.171 & 26.3 & 0.642 & 0.140 & 14.6 & Sect. 4.2, Fig.~\ref{fg:f10}\\
 D  & \citealt{kalirai08} IFMR                    & 0.742 & 0.167 & 35.5 & 0.682 & 0.142 & 20.1 & Sect. 4.3, Fig.~\ref{fg:f11}\\
 E  & \citealt{cummings16} IFMR (linear)          & 0.706 & 0.179 & 27.0 & 0.647 & 0.147 & 15.2 & Sect. 4.3, Fig.~\ref{fg:f11}\\
 F  & \citealt{catalan08} IFMR                    & 0.704 & 0.183 & 29.2 & 0.635 & 0.155 & 16.4 & Sect. 4.3, Fig.~\ref{fg:f12}\\
 G  & \citealt{casewell09} IFMR                   & 0.687 & 0.188 & 27.1 & 0.618 & 0.162 & 15.5 & Sect. 4.3, Fig.~\ref{fg:f12}\\
 H  & IMF $\alpha = 2.5$                          & 0.690 & 0.168 & 24.7 & 0.633 & 0.135 & 12.7 & Sect. 4.4, Fig.~\ref{fg:f13}\\
 I  & IMF $\alpha = 3.0$                          & 0.659 & 0.146 & 17.5 & 0.611 & 0.109 & 8.1  & Sect. 4.4, Fig.~\ref{fg:f13}\\
 J  & Constant SFH in last 12~Gyr                 & 0.685 & 0.171 & 23.8 & 0.639 & 0.141 & 14.2 & Sect. 4.5, Fig.~\ref{fg:f14}\\
 K  & \citealt{tremblay14} SFH                    & 0.717 & 0.178 & 29.8 & 0.643 & 0.139 & 13.9 & Sect. 4.5, Fig.~\ref{fg:f14}\\
 L  & No vertical scale height variation          & 0.685 & 0.168 & 23.1 & 0.632 & 0.127 & 12.5 & Sect. 4.6, Fig.~\ref{fg:f15}\\
 M  & Thick H-layers only                         & 0.701 & 0.175 & 26.8 & 0.643 & 0.143 & 14.9 & Sect. 4.7, Fig.~\ref{fg:f15}\\
 N  & \citealt{salaris10} cooling models          & 0.701 & 0.175 & 26.8 & 0.647 & 0.148 & 15.5 & Sect. 4.7, Fig.~\ref{fg:f16}\\
 O  & O/Ne-cores for $M_{\rm WD} > 1.05 M_{\odot}$& 0.701 & 0.175 & 27.0 & 0.635 & 0.130 & 13.6 & Sect. 4.7, Fig.~\ref{fg:f16}\\
 P  & Removal of magnetic WDs (SDSS only)         &  $-$  &  $-$  &  $-$ & 0.639 & 0.139 & 14.0 & Sect. 4.9, Fig.~\ref{fg:f17}\\
 \noalign{\smallskip}
 \hline
 \noalign{\smallskip}
 \multicolumn{9}{@{}p{0.85\textwidth}@{}}{{\bf Notes.} 
   The different Monte Carlo experiments ($A$ to $P$) are described
   throughout this work. The standard case $A$ includes observational errors, a Salpeter IMF ($\alpha = 2.35$), a constant SFH
   in the last 10~Gyr, the velocity dispersion versus total age relation of Eqs.~1-2, a 2nd order fit of the \citet{cummings16} IFMR
   defined in Eqs.~3-4, and the C/O-core cooling models of \citet{fontaine01}. We neglect low-mass objects
     ($M < 0.45~M_{\odot}$) for the computation of the mean masses and mass
    dispersions and this should be taken into account when comparing to other studies.
	\label{fg:t1}}
\end{tabular}
\end{table*}

\begin{table*}
%\scriptsize
\centering
\caption{SDSS Mass Distribution}
\setlength{\tabcolsep}{3.25pt}
\begin{tabular}{@{}rccccccc}
\hline
\hline
\noalign{\smallskip}
 & \multicolumn{3}{c}{Observed} & \multicolumn{3}{c}{Simulated}\\
 \multicolumn{1}{c}{Data} & $\langle M \rangle$ & $\sigma_{\rm M}$ & N$(>0.75 M_{\odot})$ & $\langle M \rangle$ & $\sigma_{\rm M}$ & N$(>0.75 M_{\odot})$\\
 \multicolumn{1}{c}{} & ($M_{\odot})$ & ($M_{\odot})$ & \% & ($M_{\odot})$ & ($M_{\odot})$ & \% \\

 \noalign{\smallskip}
 \hline
 \noalign{\smallskip}
 22,000 $< T_{\rm eff}$ (K) $<$ 30,000      & 0.599 & 0.109 &  6.4 & 0.665 & 0.166 & 20.6\\
 16,000 $< T_{\rm eff}$ (K) $<$ 22,000      & 0.619 & 0.108 & 10.0 & 0.642 & 0.142 & 14.6\\
 12,000 $< T_{\rm eff}$ (K) $<$ 16,000      & 0.643 & 0.100 & 10.8 & 0.647 & 0.139 & 16.1\\
 8000   $< T_{\rm eff}$ (K) $<$ 12,000      & 0.636 & 0.106 & 12.5 & 0.649 & 0.129 & 16.8\\
 \noalign{\smallskip}
 \hline
 \noalign{\smallskip}
 \multicolumn{8}{@{}p{0.65\textwidth}@{}}{{\bf Notes.}
   The observations include DA and DB white dwarfs for $T_{\rm eff} >$ 16,000 K
   but only DAs for smaller temperatures. 
 \label{fg:t2}}
\end{tabular}
\end{table*}

\subsection{Selection Effects}

\begin{figure}
  \centering 
  \includegraphics[scale=0.475,bb=30 147 572 589]{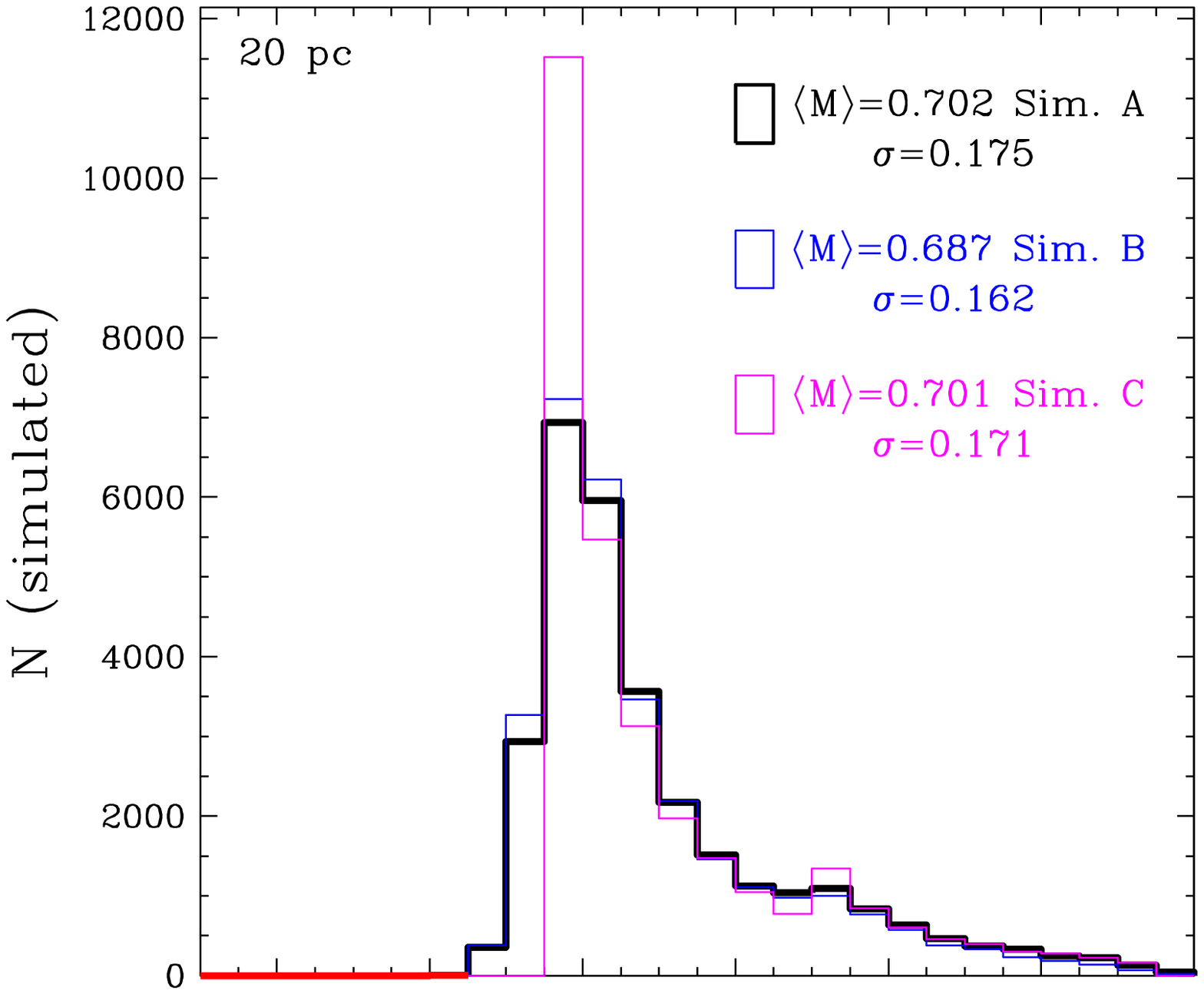}
  \includegraphics[scale=0.475,bb=30 147 572 539]{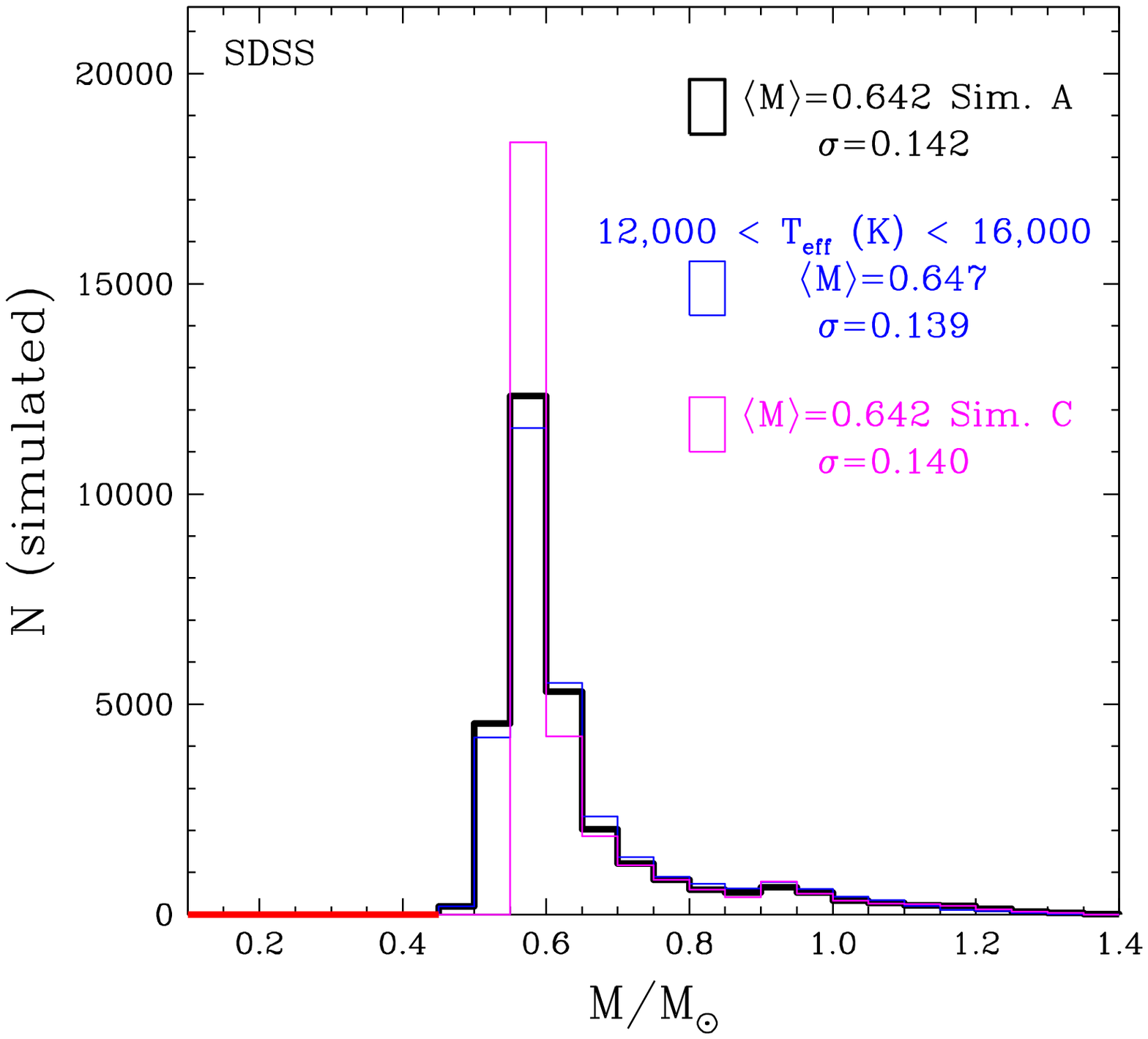}
  \caption[f10.eps]{Comparison of our standard Monte Carlo simulations $A$
    (thick black lines) with alternative experiments for the 20~pc (top panel)
    and SDSS (bottom) samples. Case $B$ is with a faint magnitude limit of
    $V<17$ (20~pc sample only) and case $C$ neglects observational mass
    errors in both panels. For the SDSS sample, we show the distribution for 12,000 $<
    T_{\rm eff}$ (K) $<$ 16,000, which contrasts with our standard case in the
    range 16,000 $< T_{\rm eff}$ (K) $<$ 22,000. There are 30,000 simulated objects
		for both samples. 
    \label{fg:f10}}
\end{figure}

This section discusses completeness issues regarding single white dwarfs,
while effects from missing white dwarfs in binaries are reviewed in Section
4.8.  The 20~pc sample is only 80-90\% complete, hence missing objects could
impact the mass distribution if they tend to be fainter and more massive than
the average. To test this scenario, we have a added a magnitude cutoff of $V <
17$ to our simulation $B_{\rm 20pc}$ in Fig.~\ref{fg:f10}.  This removes
nearly 7\% of the sample and brings the mean mass down by about 0.01
$M_{\odot}$ according to Table~\ref{fg:t1}. It confirms that objects at the
faint end are significantly more massive than the average. Yet, this magnitude
cutoff is largely insufficient to bring the fraction of high-mass white dwarfs
in agreement with the observations. \textit{Gaia} will substantially improve
the completeness by detecting most white dwarfs in the local sample down to $V
\sim 20$.

One interesting finding of this work is that about 2\% of all local degenerate
stars, corresponding to about 2 objects within 20~pc, are massive white dwarfs
within the so-called ultra-cool regime, i.e. with temperatures well below
4000~K. For $M_{\rm WD} \sim 1.0$ $M_{\odot}$, a star that formed 10~Gyr ago
spent a negligible amount of time on the main-sequence and is now a 4000~K
white dwarf. As the mass further increases, it takes less and less time to
cool to 4000~K. For $M_{\rm WD} = 1.2$ $M_{\odot}$, all stars formed more than
7~Gyr ago are now massive ultracool white dwarfs, some of them with
temperatures well below 2000~K. We find that this interpretation remains valid
even when employing alternative cooling models (see Section~4.7), including
O/Ne-core cooling tracks. While not of immediate concern for this work, it
will become an important issue for the definition of halo white dwarf samples
in the \textit{Gaia} and \textit{Euclid} era.

We have already studied the selection effects for the SDSS sample in
Section~2.  Another way to confirm our results is to select different $T_{\rm
  eff}$ subsamples and compare to our standard case. Table~\ref{fg:t2}
presents the comparison between observations and simulations for different
$T_{\rm eff}$ regimes. From experiments similar to the one presented in
Fig.~\ref{fg:f3}, we have verified that the spectroscopic completeness does
not change significantly as a function of mass over the $T_{\rm eff}$ ranges
identified in Table~\ref{fg:t2}. However, the completeness as a function of
$T_{\rm eff}$ is not constant. Furthermore, masses for non-DA white dwarfs are
not available for $T_{\rm eff} < 16,000$~K. Nevertheless, Table~\ref{fg:t2}
demonstrates that the high-mass fraction is overpredicted at all temperatures.

Fig.~\ref{fg:f10} shows that the predicted SDSS mass distribution in the range
$12,000 < T_{\rm eff}$ (K) $< 16,000$ is fairly similar to our warmer standard
case.  It is difficult to predict mean mass variations as a function of
$T_{\rm eff}$ since it depends significantly on the variation of the vertical
scale height as a function of total stellar age. Indeed, distances for the
magnitude-limited SDSS sample are strongly correlated with $T_{\rm eff}$
values. As a consequence, the variations in the simulated mean masses
presented in Table~\ref{fg:t2} should be taken as indicative only.

The observations also show mean mass fluctuations as a function of temperature
according to Table~\ref{fg:t2}.  They are thought to be caused by
spectroscopic calibration issues \citep{kleinman04,tremblay11}, and to a
lesser degree missing subtypes as well as incomplete 3D effects
\citep{tremblay13}. The data calibration issues were first discussed for the
SDSS DR1 sample \citep{kleinman04} and the status appears largely unchanged in
our analysis employing the DR10 reduction. In the $16,000 < T_{\rm eff}$ (K)
$< 22,000$ regime of our standard sample, the SDSS spectra lead to lower
derived masses compared to independent observations \citep[see,
e.g.,][]{gianninas11}. An account of this bias would generate a better
agreement between the observations and our standard SDSS simulation.

\subsection{Observational Errors}

Most white dwarfs are formed close to the $\sim$0.6 $M_{\odot}$ peak and the
observed shape of that peak is largely determined by how it is convolved with
observational errors. This effect is confirmed by Case $C$ in
Fig.~\ref{fg:f10} where we have removed observational errors.  However, there
is only a small impact on the mean mass, the fraction of high-mass objects,
and even the mass dispersion. Nevertheless, it demonstrates that it would be
necessary to perform a more careful assessment of the observational errors,
possibly including asymmetries, to properly fit the observed white dwarf mass
distributions with a grid of Monte Carlo simulations.

\subsection{Initial-Final Mass Relation}

The IFMR relation is clearly a critical parameter to map the IMF into the
white dwarf mass distribution \citep{catalan08}. While the intermediate-mass
IFMR is relatively well understood \citep[see, e.g.,][]{cummings15}, the slope
at the low-mass end, roughly defined as $M_{\rm WD} < 0.65$ $M_{\odot}$
($M_{\rm i} \lesssim 2.5$ $M_{\odot}$), is still poorly constrained
\citep{catalan08,kalirai08,kalirai09,zhao12}. This has a crucial impact on the
simulated mass distributions since the mass peak is well within this
regime. Furthermore, the high-mass end of the IMFR ($M_{\rm WD} > 1.0$
$M_{\odot}$) is also poorly explored since massive white dwarfs are rare in
clusters \citep{williams09,cummings16,cummings16b,raddi16}, and one has to
rely on an extrapolation to predict the high-mass tail of the simulated mass
distributions.

\begin{figure}
  \centering 
  \includegraphics[scale=0.475,bb=30 147 572 589]{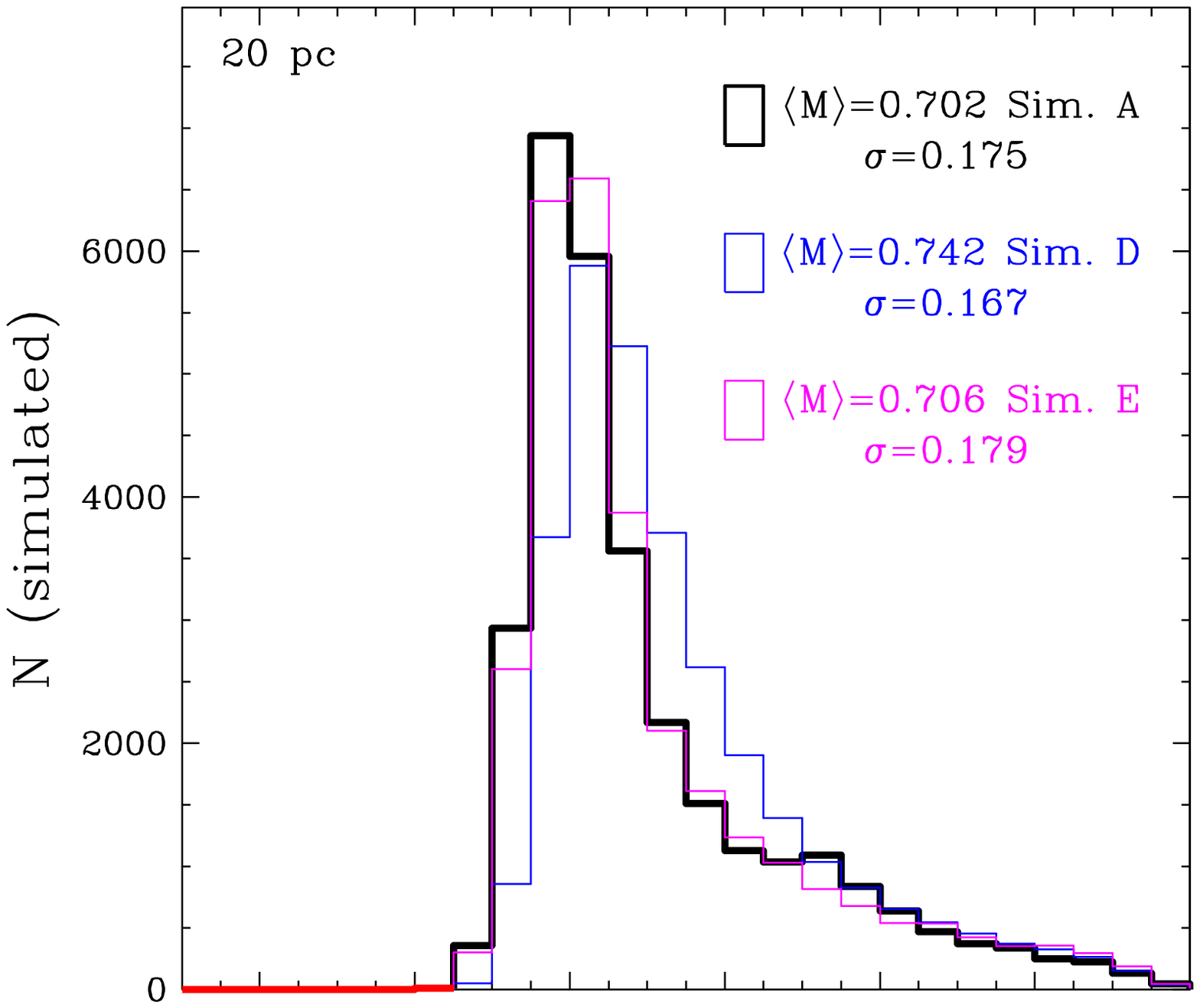}
  \includegraphics[scale=0.475,bb=30 147 572 539]{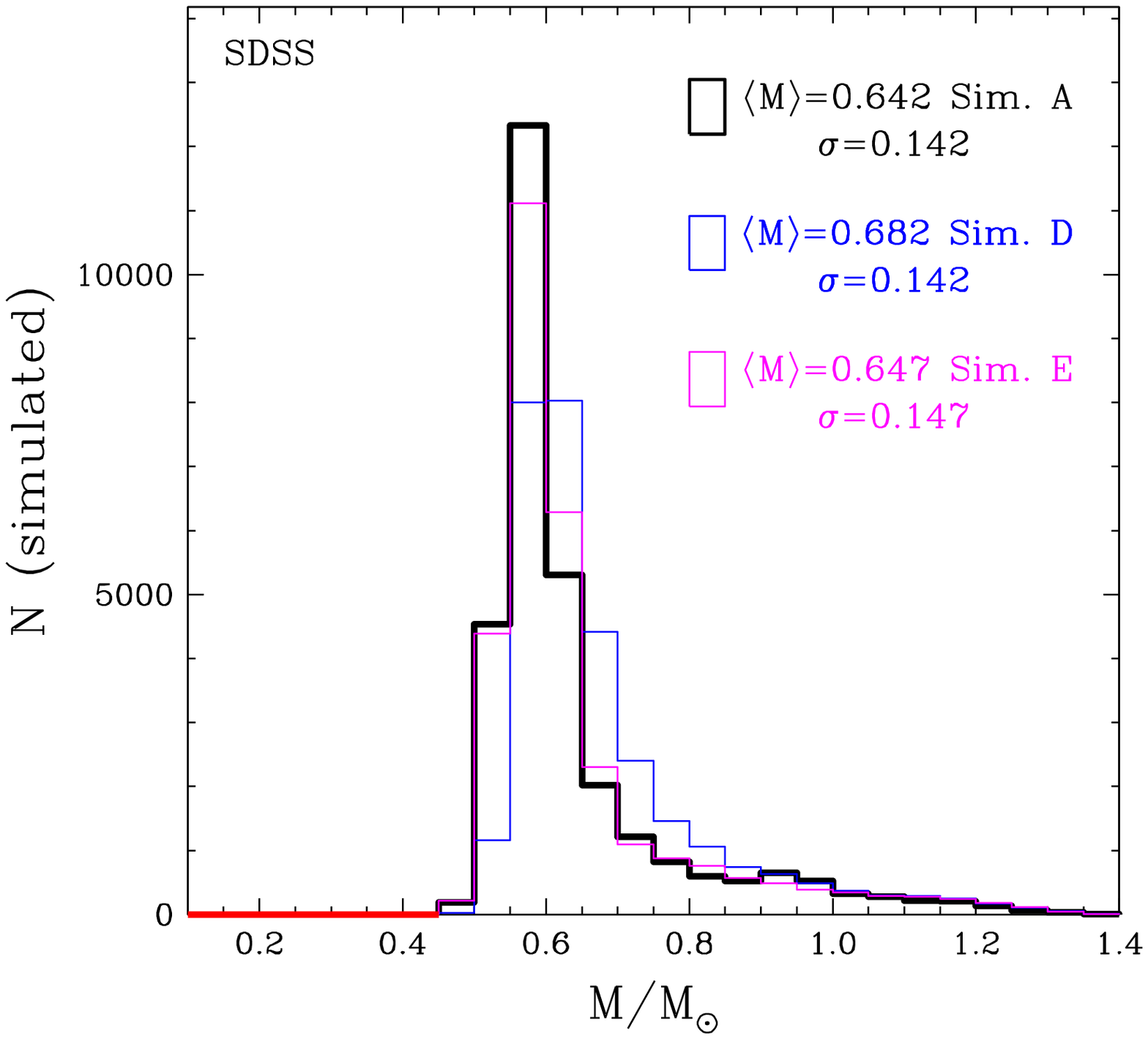}
  \caption[f11.eps]{Similar to Fig.~\ref{fg:f10} but with alternative
    numerical experiments. Case $D$ employs the IFMR of \citet{kalirai08} and
      case $E$ uses a linear fit to the \citet{cummings16} IFMR. For the
      standard case $A$ we rely on a two-part 2nd order fit to the
      \citet{cummings16} IFMR.
 \label{fg:f11}}
\end{figure}

\begin{figure}
  \centering 
  \includegraphics[scale=0.475,bb=30 147 572 589]{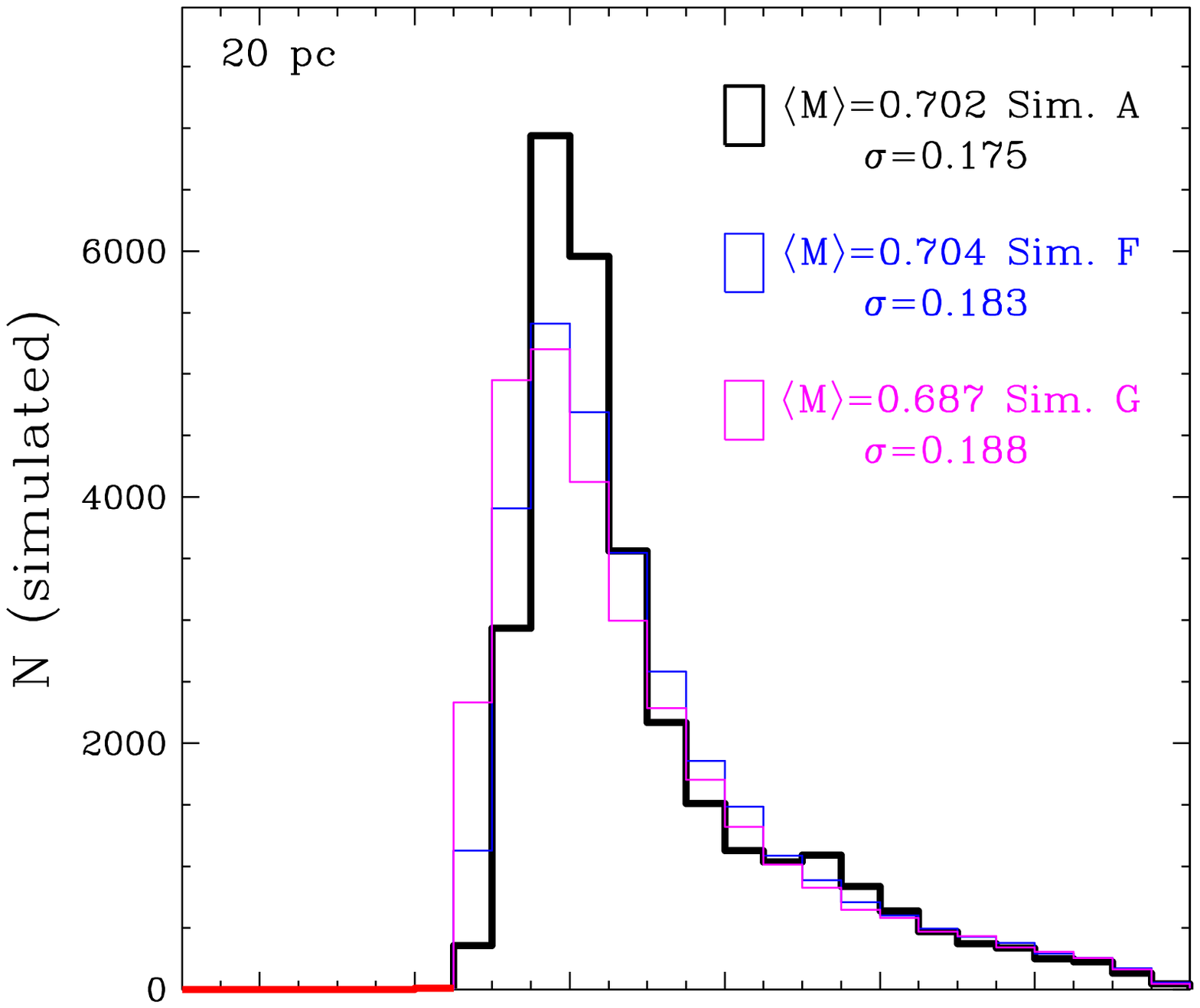}
  \includegraphics[scale=0.475,bb=30 147 572 539]{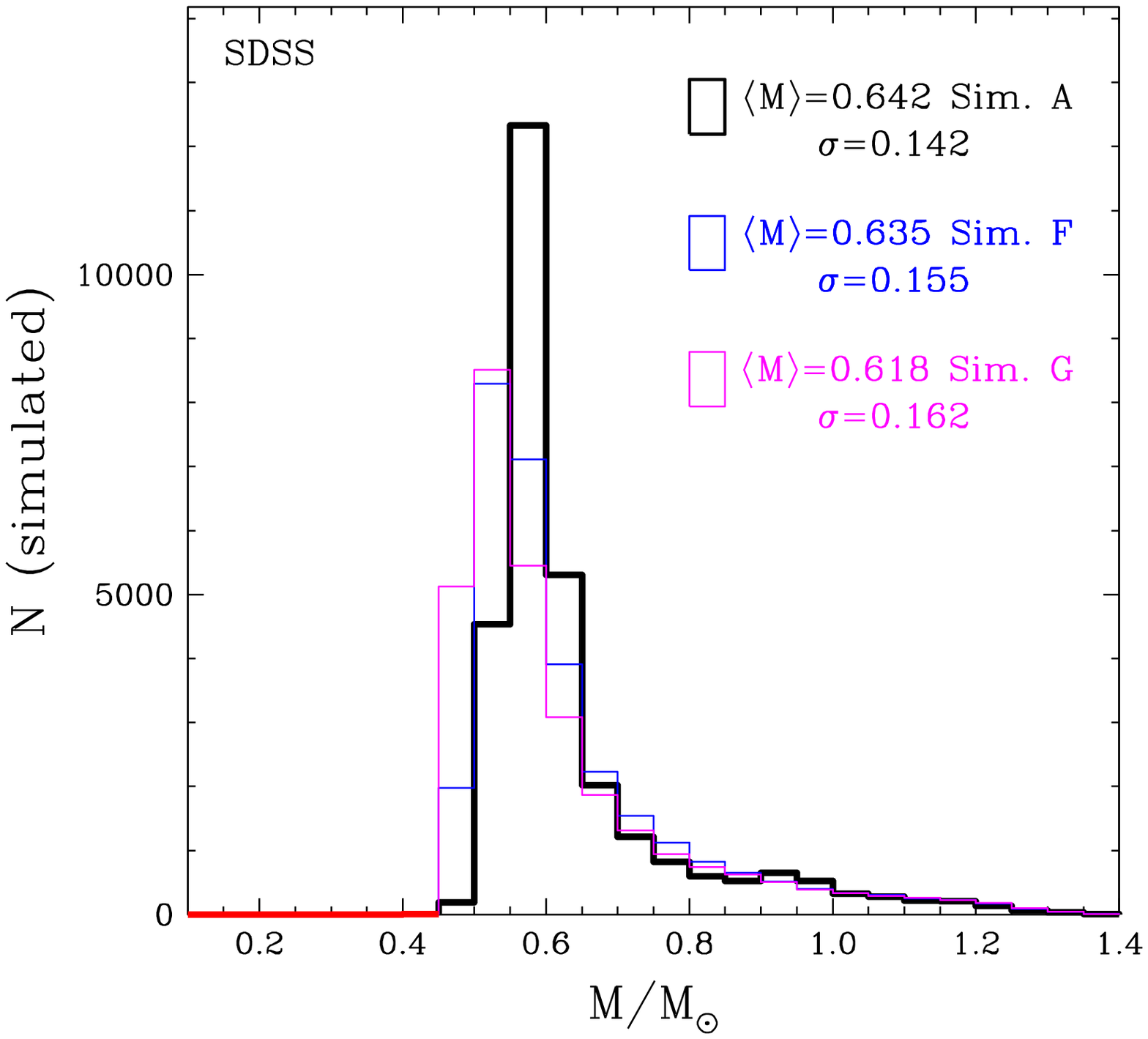}
  \caption[f12.eps]{Similar to Fig.~\ref{fg:f10} but with alternative
    numerical experiments. Case $F$ relies on the IFMR of \citet{catalan08} and
      case $G$ uses the relation from \citet{casewell09}. 
 \label{fg:f12}}
\end{figure}

Figs.~\ref{fg:f11} and \ref{fg:f12} present our results with a set of four
alternative IFMRs, all of them linear relations. For cases $D$, $E$, $F$, and
$G$, respectively, we employ the relation of \citet{kalirai08}, the results of
\citet{cummings16} as used in our standard case but with a linear instead of a
two-part 2nd order fit, the parameterisation of \citet{catalan08}, and the
IFMR from \citet{casewell09}. As expected, these alternative assumptions have
a strong impact on the predicted mass distributions. For both samples, the
mean mass varies by as much as $\sim$0.06 $M_{\odot}$, while the fraction of
massive white dwarfs changes by up to 8\%. It is therefore clear that the low-
and high-mass regimes of the IFMR must be better understood to predict the
field white dwarf mass distribution. However, all of our assumed IFMRs predict
a too large amount of massive white dwarfs, suggesting it is unlikely to be
the only source of the discrepancy.

\subsection{Initial Mass Function}

\begin{figure}
  \centering 
  \includegraphics[scale=0.475,bb=30 147 572 589]{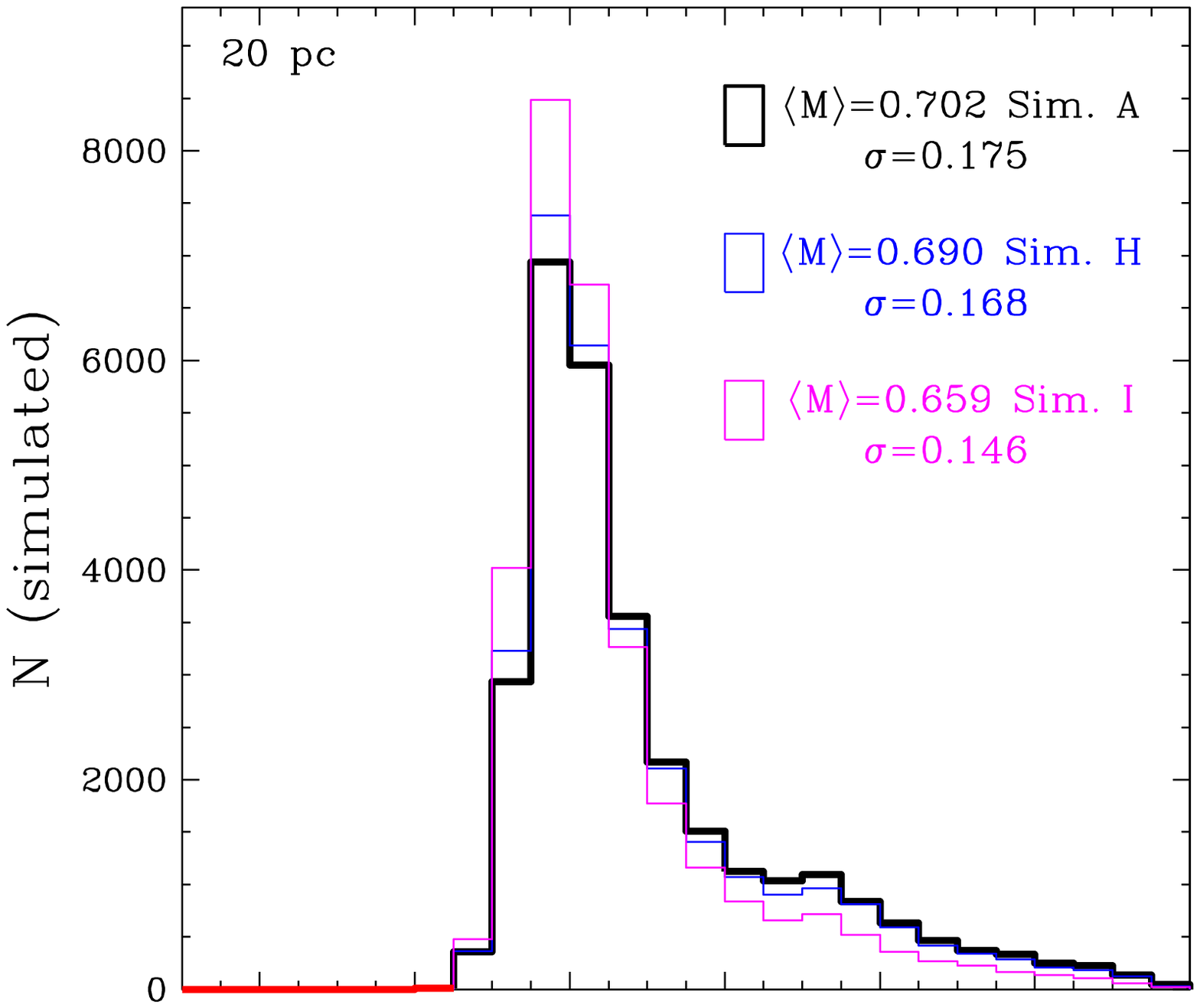}
  \includegraphics[scale=0.475,bb=30 147 572 539]{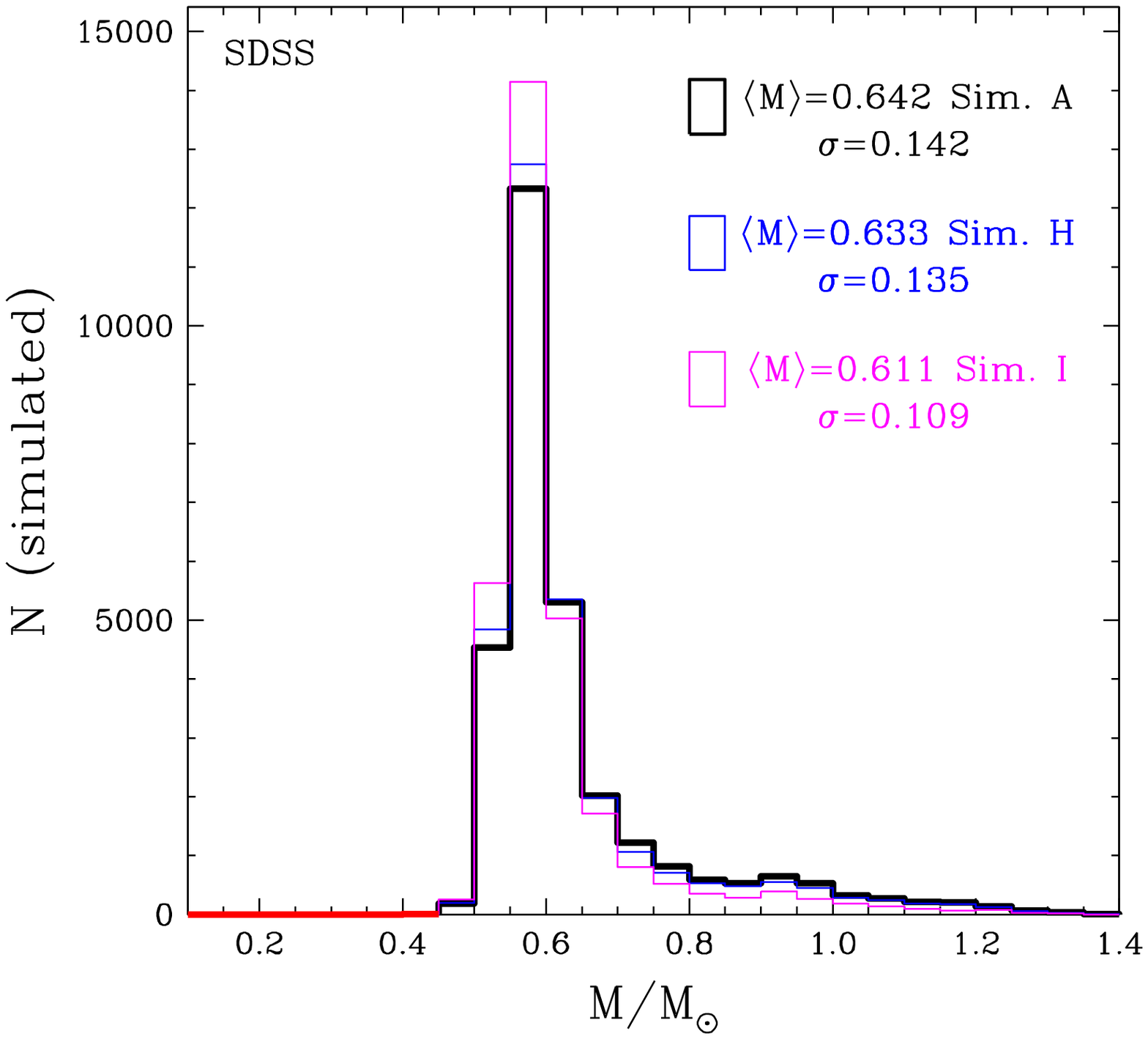}
  \caption[f13.eps]{Similar to Fig.~\ref{fg:f10} but with alternative
    numerical experiments. We employ an IMF with a power index of $\alpha = 2.5$
for case $H$ and $\alpha = 3.0$ for case $I$, while the standard case uses
the Salpeter value of $\alpha = 2.35$.
\label{fg:f13}}
\end{figure}

We have repeated our simulations with a steeper IMF. Instead of the Salpeter
relation, we have employed $\alpha$ = 2.5 and 3.0 in cases $H$ and $I$,
respectively. The results are shown in Fig.~\ref{fg:f13}. For both samples,
Table~\ref{fg:t1} suggests that an IMF slightly steeper than $\alpha$ = 2.5
would put the mean mass, mass dispersion, and massive white dwarf fraction in
fairly good agreement with the observations. We note that \citet{imf} find a
high-mass IMF of $\alpha$ = 2.45$^{+0.03}_{-0.06}$ from young clusters in M31.
Furthermore, \citet{bochanski10} have used SDSS data to derive a low-mass
single star IMF (0.32 $< M_{\rm i}/M_{\odot} < 0.8$) that is consistent with
$\alpha \sim$ 2.60.  Our results at face value also suggest a single star IMF
steeper than Salpeter for the disk of the Milky Way in the range 1.0 $< M_{\rm
  i}/M_{\odot} < 8.0$. It is however difficult to isolate the effect of the
IMF from other input parameters.

\subsection{Stellar Formation History}

\begin{figure}
  \centering 
  \includegraphics[scale=0.475,bb=30 147 572 589]{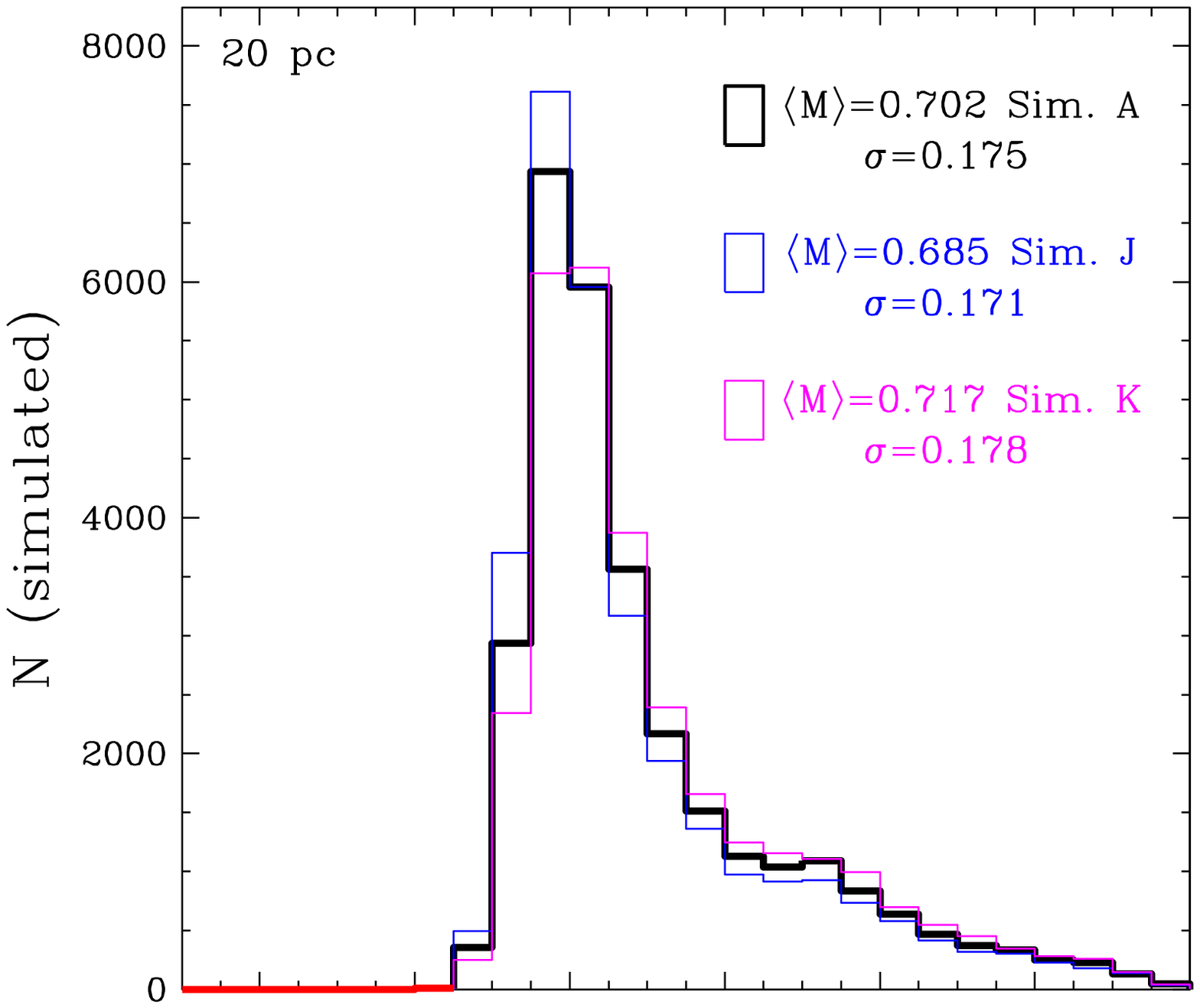}
  \includegraphics[scale=0.475,bb=30 147 572 539]{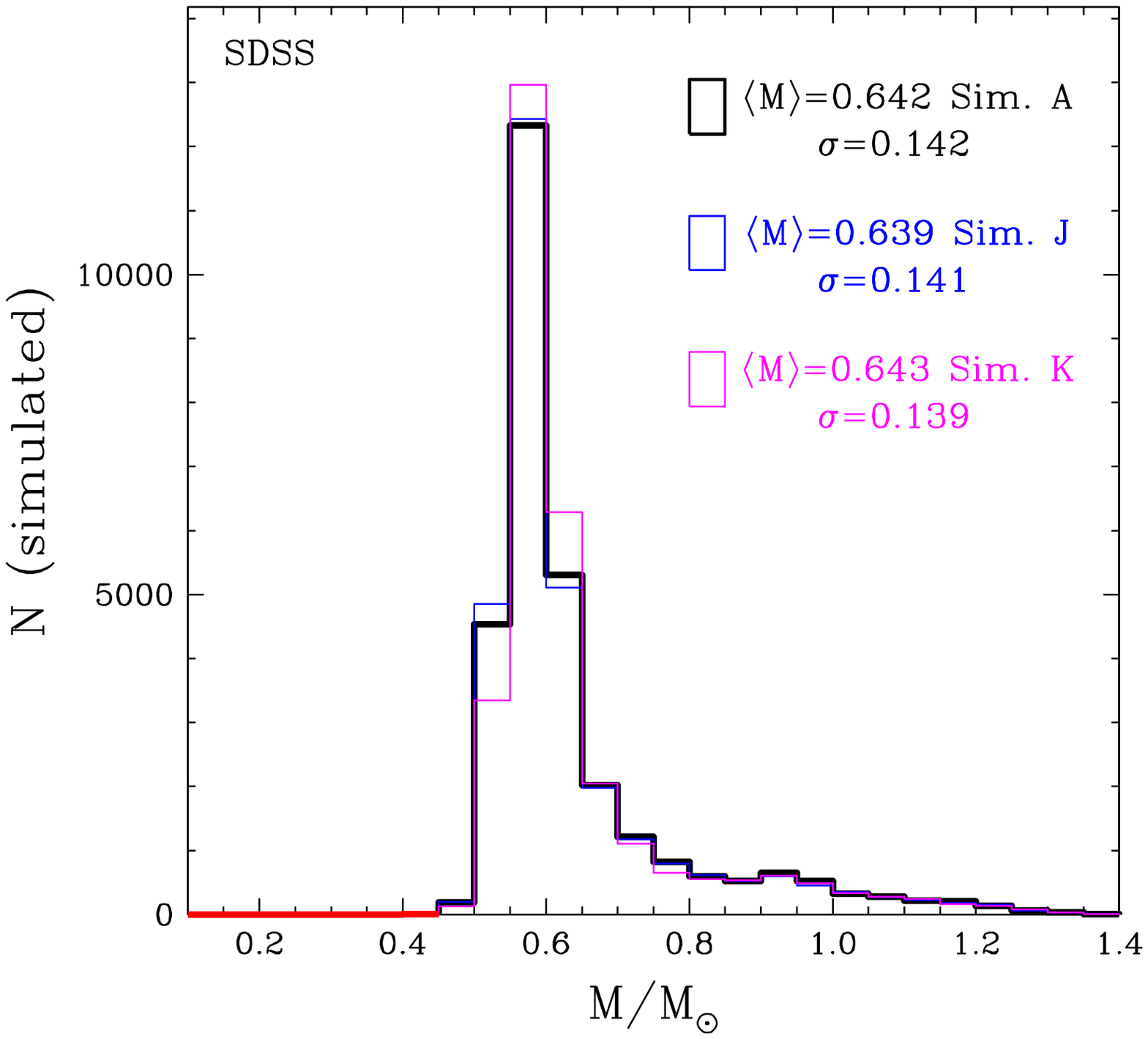}
  \caption[f14.eps]{Similar to Fig.~\ref{fg:f10} but with alternative
    numerical experiments. For case $J$ we employ an age of 12~Gyr instead of
    10~Gyr for the Galactic disk and case $K$ uses the SFH derived in
    \citet{tremblay14} instead of a constant formation rate.
 \label{fg:f14}}
\end{figure}

Our standard case assumes a constant SFH for the Galactic disk in the last
10~Gyr. Fig.~\ref{fg:f14} presents the results supposing instead an age of
12~Gyr for the disk. The effect is quite important for the 20~pc sample as it
greatly enhances the number of $\sim$1 $M_{\odot}$ stars that became white
dwarfs. The effect on the SDSS mass distribution of young white dwarfs is much
smaller. While it is clear that our experiment overestimates the age of the
disk \citep[see, e.g.,][]{winget87}, it illustrates that one has to obtain a
precise estimate of this parameter to model the field white dwarf mass
distribution.

We have recently constrained the local SFH from white dwarfs within 20~pc
\citep{tremblay14}. We have not used this result so far since the technique
employed to derive the SFH is more sensitive at intermediate ages ($\sim$3-10
Gyr) and constraints on the last 2~Gyr depend much more on the assumed IMF.
In particular, for the SDSS sample where most massive white dwarfs are from
stars formed in the last 1~Gyr, it is difficult to apply our earlier SFH
results. Nevertheless, Fig.~\ref{fg:f14} presents the case $K$ where we have
used the white dwarf determined SFH from \citet{tremblay14} instead of a
constant value. The impact on the mass distributions is moderate. Our input
SFH peaks at 2-4 Gyr and Fig.~\ref{fg:f6} shows that no massive SDSS white
dwarf is found in that range, resulting in a smaller high-mass fraction for
that sample. It is the opposite situation for the 20~pc sample where the
high-mass fraction increases. There is currently no consensus on the SFH and
radial migration within the Galactic disk \citep[see, e.g.,][]{tremblay14}.
As a consequence, it is difficult to quantify the amplitude and sign of this
bias.

\subsection{Vertical Scale Height of the Galactic Disk}

Previous studies of white dwarf luminosity functions have 
often assumed a constant vertical scale height of 250~pc for the Galactic disk
\citep[see, e.g.,][]{harris06, torres16}. It is nonetheless known
that cooler white dwarfs have a larger vertical ($W$) velocity dispersion. 
From the \citet{sion14} kinematic analysis of the 25~pc sample, it is possible
to divide the sample for cooling ages below and above 1.37 Gyr ($T_{\rm eff} \sim 8000$~K)
and conclude that the older bin has a larger vertical velocity dispersion by a factor of $\sim$1.5.
If we do the same analysis for our standard local sample simulation,
we find a ratio of 1.4-1.9 depending on the Galactic disk model used to
transform scale height into velocity dispersion. This suggests that our scale height 
variation model drawn from main-sequence star observations is appropriate. 
Nevertheless, there are very few studies that constrain
the absolute values of the vertical scale height of white
dwarfs and one should be cautious with the predictions of our standard
simulations.

\begin{figure}
  \centering 
  \includegraphics[scale=0.475,bb=30 147 572 589]{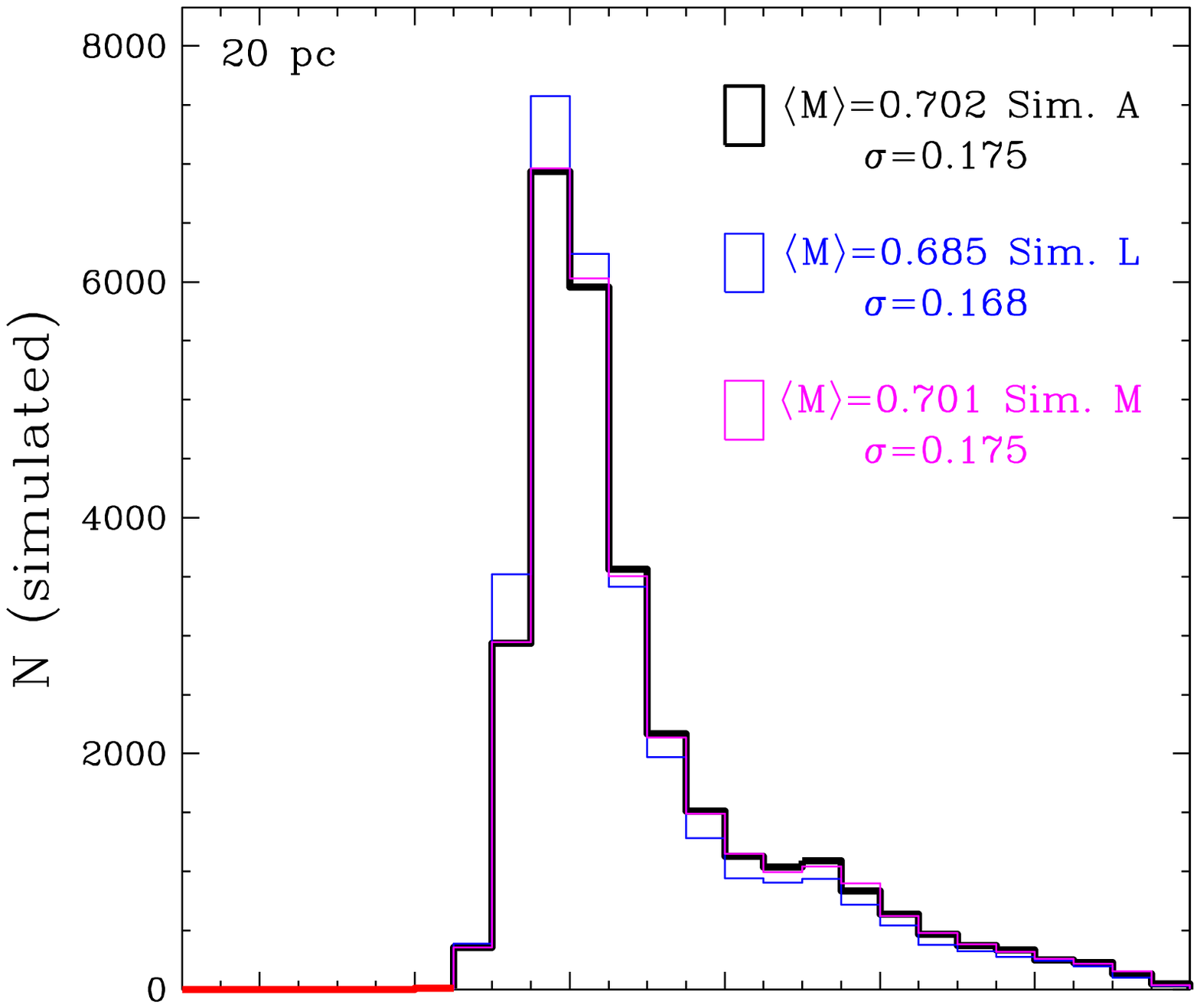}
  \includegraphics[scale=0.475,bb=30 147 572 539]{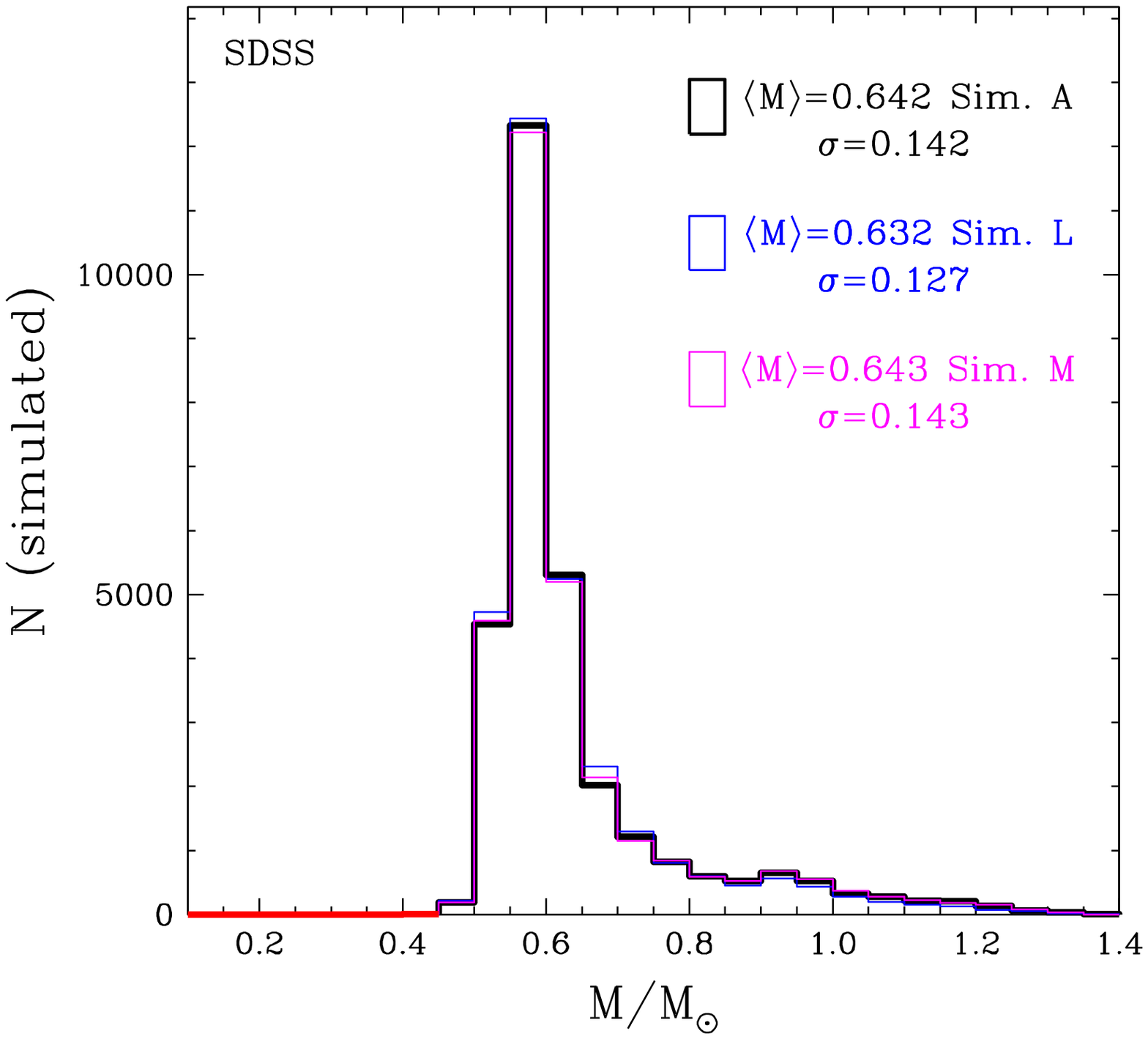}
  \caption[f15.eps]{Similar to Fig.~\ref{fg:f10} but with alternative
    numerical experiments. We assume that the vertical scale height of the
    Galactic disk has a constant value of 250~pc in case $L$ and that all white dwarfs
    have thick H-layers instead of 70\% with thick and 30\% with thin H-layers
    in case $M$.
    \label{fg:f15}}
\end{figure}

Fig.~\ref{fg:f15} shows the case $L$ where we assume a constant vertical scale
height of 250~pc for the Galactic disk. This alternative parameterisation has
significant consequences since high-mass white dwarfs are now formed at much
higher Galactic latitudes on average. For both the SDSS and 20~pc samples,
Fig.~\ref{fg:f6} shows that most massive degenerates are detected close to the
plane of disk where the Sun is located. This reduces the simulated fractions
of massive white dwarfs as seen in Table~\ref{fg:t1}, in better agreement with the 
observations. Lower mass white dwarfs
are relatively unaffected since they already have a scale height of
$\sim$200~pc in our standard simulations owing to their large total ages on
average. Nevertheless, it appears unrealistic that the vertical scale
height is constant or decrease with time, hence it is unlikely that it is the
main reason for the overprediction of massive white dwarfs.

\subsection{White Dwarf Models and Evolution Tracks}

\begin{figure}
  \centering 
  \includegraphics[scale=0.475,bb=30 147 572 589]{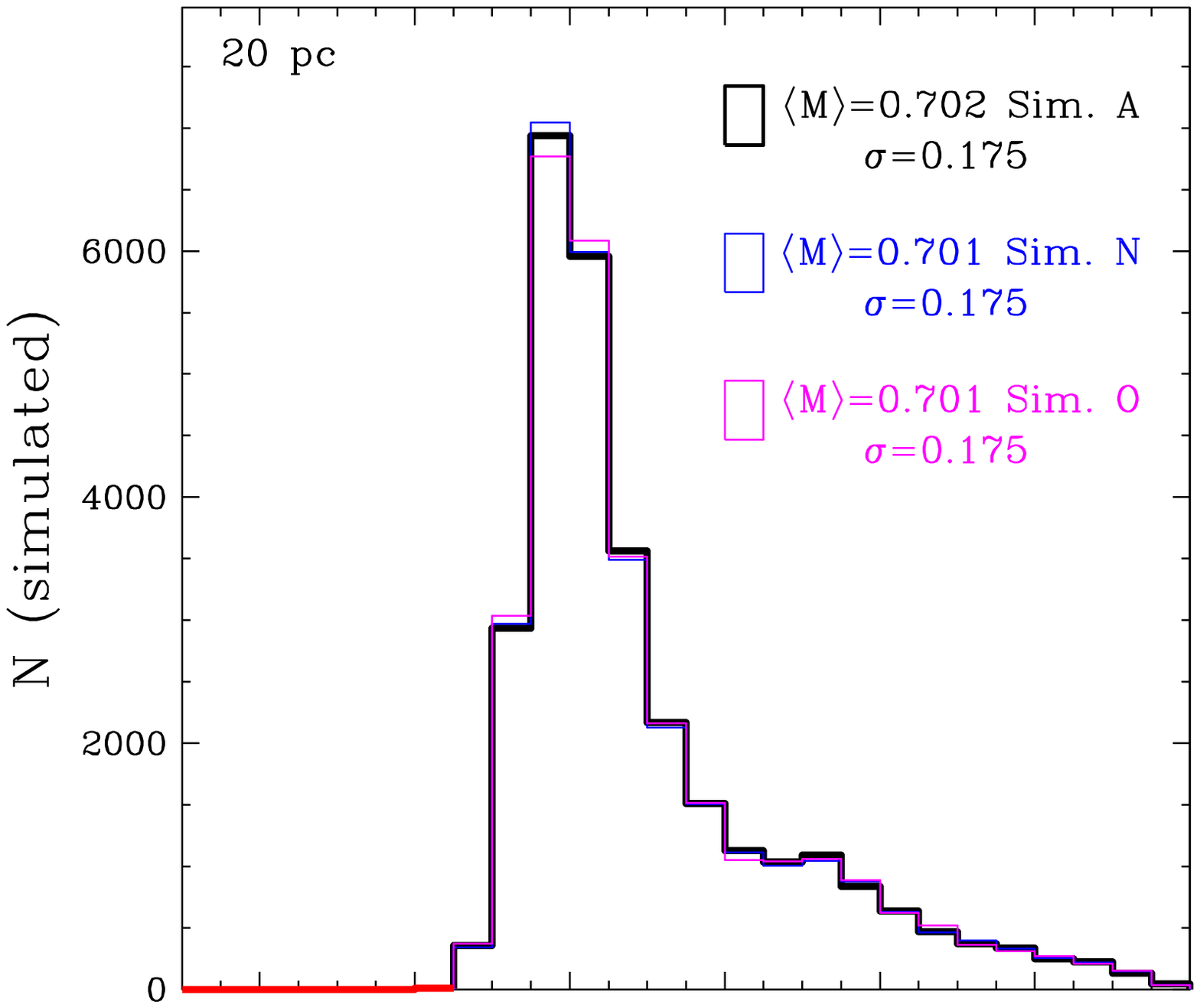}
  \includegraphics[scale=0.475,bb=30 147 572 539]{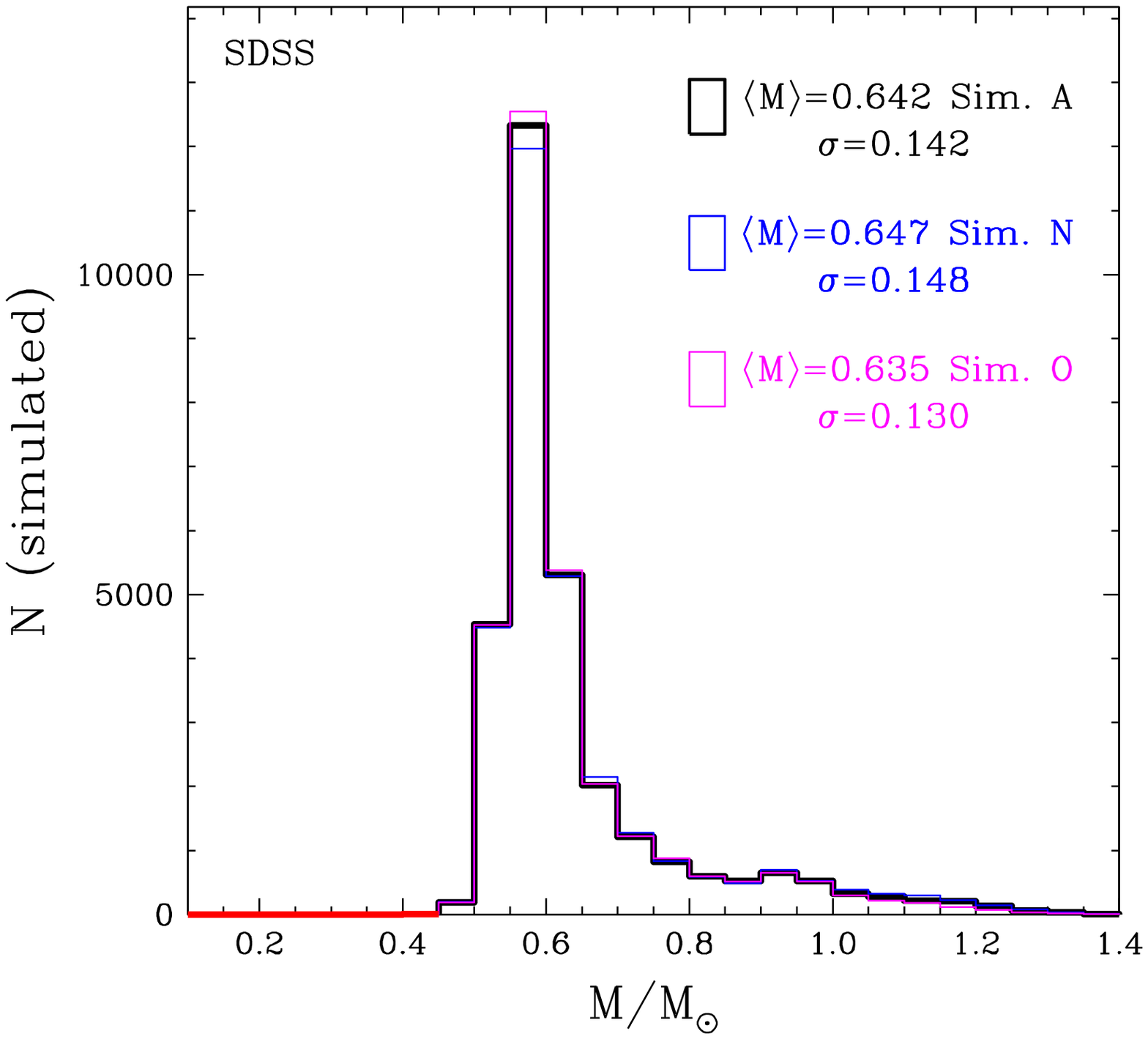}
  \caption[f16.eps]{Similar to Fig.~\ref{fg:f10} but with alternative
    numerical experiments. We employ the cooling sequences of
    \citet{salaris10} for C/O-core white dwarfs in case $N$ and O/Ne-core
    evolutionary sequences from \citet{althaus07} for $M_{\rm WD} > 1.05$~$M_{\odot}$ in
    case $O$.
    \label{fg:f16}}
\end{figure}

For case $M$, we have used thick hydrogen layers for all objects and
Fig.~\ref{fg:f15} demonstrates that the effect is negligible compared to other
biases. Additionally, we have employed alternative cooling sequences from
\citet{salaris10} in case $N$, where effects of C/O phase separation and
sedimentation are taken into account. We have also used \citet{althaus07}
evolutionary sequences with O/Ne cores for $M_{\rm WD} > 1.05$ $M_{\odot}$ in
case $O$. Fig.~\ref{fg:f16} demonstrates that changes are small for both
experiments. Regarding the high-mass fraction, the effects on the 20~pc sample
are negligible since the cooling rates do not change the distance or
membership. For the SDSS sample, we note that O/Ne cores reduce the number of
high-mass white dwarfs, in the direction of bringing the simulations in better
agreement with the observations.

Further uncertainties lie in the model atmospheres and fitting techniques used
to extract the observed mass distributions. For instance, Fig.~\ref{fg:f2}
shows the overall SDSS mass distribution for the same DR7 sample of DA white
dwarfs, but with spectroscopic masses determined by two independent studies.
The differences are moderate, and most often not significant for a single
spectrum, but still lead to systematic effects on the mean mass and high-mass
fraction.

\subsection{Binaries}

We have so far neglected unresolved binaries both in our simulations and
observed distributions. That includes WD+MS and WD+WD binaries, where WD
stands for white dwarf and MS for main-sequence. We discuss merger products
separately in Section~5.1.

The fraction of WD+MS binaries as function of initial mass is likely to vary
strongly and binarity appears to be more common in massive stars
\citep{binary3}, hence high-mass white dwarfs. Such scenario is difficult to
constrain from white dwarf populations because we have little information on
the unbiased mass distribution of white dwarfs in binaries. Increasing the
fraction of binaries as a function of initial mass would be similar to using a
steeper single star IMF for our single white dwarf simulations. This would
make our simulations in better agreement with the observations.

Our cutoff below 0.45 $M_{\odot}$ should eliminate most He-core white dwarfs
formed through binary evolution. However, post-common envelope binaries also
include C/O-core white dwarfs \citep[see, e.g.,][]{alberto11,binary1}. Some of
these systems are likely present in our sample when the companion is an unseen
low-mass star or a white dwarf.  In those cases, the brighter and lower mass
white dwarf will likely have suffered mass loss.  Accounting for this effect
would bring our simulations in closer agreement with the observations.

Finally, double degenerates that have not previously interacted could also be
a problem since the lower mass white dwarf will dominate the flux and massive
white dwarf companions could be hidden. \citet{tremblay11} find a $\sim$1\%
fraction of DA+DB/DC double degenerates in the SDSS, suggesting a five times
larger DA+DA fraction given that the ratio of hydrogen to helium atmospheres
is about 5/1 \citep{kleinman13}. Only a small number of those are expected to
have large mass ratios, suggesting that double degenerates may not
significantly impact the observed mass distributions.

\subsection{Magnetic White Dwarfs}

\begin{figure}
  \centering 
  \includegraphics[scale=0.475,bb=30 147 572 589]{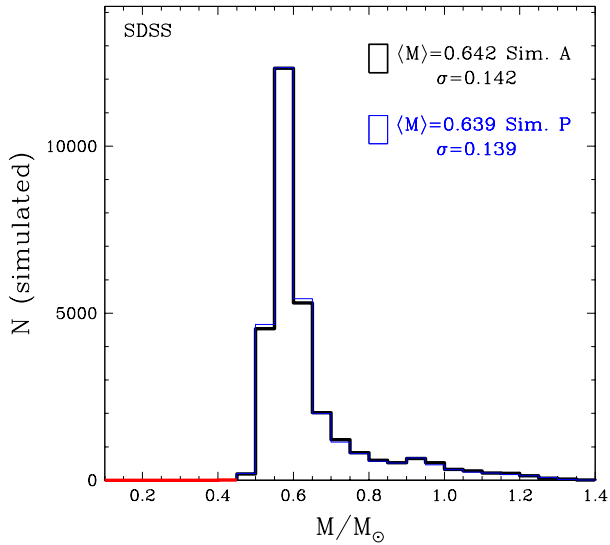}
  \caption[f17.eps]{Similar to Fig.~\ref{fg:f10} but for the alternative
    numerical experiment $P$ (SDSS only) where we have removed
    the contribution from a population of magnetic white dwarfs with 
    an incidence of 2.5\% and a mean mass of 0.80 $M_{\odot}$.
    \label{fg:f17}}
\end{figure}

Magnetic white dwarfs in the 20~pc sample are included in the observed mass
distribution of Fig.~\ref{fg:f1} because they have precise masses from
trigonometric parallax measurements. On the other hand, we have neglected
magnetic white dwarfs from the observed SDSS distribution since there are no
mass estimates for them. We have identified 2.5\% of magnetic DA white dwarfs
in our revised analysis of the SDSS DR7 sample. We can account for these
objects by assuming that the same fraction of our simulated white dwarfs are
magnetic. It is suggested that the mass distribution for magnetic degenerates
peaks around $\sim$0.8 $M_{\odot}$ \citep{briggs15,ferrario15}.  We note that
the 15 magnetic degenerates in our 20~pc sample have a mean mass of 0.75
$M_{\odot}$, which is 11\% larger than the non-magnetic white dwarfs.  As a
consequence, we assume that the probability of an object being magnetic varies
linearly with mass. The slope and amplitude of this function are fixed to
obtain a mean mass of 0.8 $M_{\odot}$ and a magnetic fraction of 2.5\%.
Finally, we have removed those magnetic objects in case $P$ presented in
Fig.\ref{fg:f17}. The impact is to reduce the mean mass and high-mass
fraction, though the effect is moderate given the small total number of
magnetic white dwarfs. We mention that if these magnetic white dwarfs come
from mergers, our simulations do not adequately represent them.

\section{Discussion}

The white dwarf mass distributions from the 20~pc and SDSS samples were first
studied by designing Monte Carlo simulations with fixed standard astrophysical
constraints.  The good qualitative agreement between simulations and
observations in Figs.~\ref{fg:f1} and \ref{fg:f4} confirms that the local
white dwarf population is consistent with our basic knowledge of stellar and
Galactic evolution.  We have then systematically studied the uncertainties on
the input parameters of the simulations. Our results suggest that given our
current knowledge of stellar and Galactic evolution, we can only predict the
mean mass and mass dispersion of observed samples within
$\sim$10\%. Additionally, we find that the relative number of high-mass white
dwarfs ($M > 0.75$ $M_{\odot}$) can only be predicted within a factor of
$\sim$1.5. The main uncertainties are the assumed IFMR and the IMF, followed
by the SFH, the scale height variation of the Galactic disk as a function
total stellar age, and binaries. Other biases lead to moderate changes, such
as effects from missing magnetic white dwarfs (SDSS sample), incompleteness
(20~pc sample), white dwarf model atmospheres and evolution tracks, and core
composition. Finally, we find that observational errors lead to fairly small
uncertainties on the mean properties and high-mass fraction.

From the same local white dwarf sample as the one studied in this work,
\citet{tremblay14} have successfully extracted the local SFH in the last
10~Gyr. We have verified that if we observe our Monte Carlo simulation $K_{\rm
  20pc}$ with added observational errors, we can recover the input SFH from
\citet{tremblay14} to a high precision with the technique described in that
work. Our present study does not lessen the significance of this recent
determination of the SFH in the solar neighborhood. We demonstrate instead
that it is difficult to extract the IMF from the same sample. The first reason
for this behaviour is that \citet{tremblay14} have used a direct method
employing both the mass and cooling age of individual white dwarfs. In the
present case, we consider the mass distribution integrated over all ages.  The
second reason why it is difficult to constrain the IMF is that the IMFR leads
to similar effects on the mass distribution. Finally, biases owing to binary
populations and incompleteness directly impact the mass distributions, while
\citet{tremblay14} have demonstrated that it does not lead to significant
systematic effects on the SFH.

We find that our simulations overpredict the fraction of massive white dwarfs
by a factor of $\sim$1.5 for both the 20~pc and SDSS samples. This
interpretation is consistent with earlier Monte Carlo simulations of similar
populations \citep[see, e.g.,][]{catalan08}. This result suggests a single
star IMF that is significantly steeper than Salpeter for the Galactic disk.
However, if we account for all uncertainties, a Salpeter IMF is not ruled
out. We note that this result differs from the common view that there is an
observed excess of massive white dwarfs when representing the peak in the mass
distributions with a Gaussian function \citep[see, e.g.,][]{kleinman13}. We do
not challenge this fact but only the astrophysical interpretation. Our
calculations suggest that Gaussian functions are a poor substitute to
realistic simulations including stellar and Galactic evolution when attempting
to understand the nature of high-mass white dwarfs.

\textit{Gaia} will soon provide precise parallaxes for all white dwarfs
studied in this work. This will supply precise independent masses leading to a
better understanding of the observed mass distributions. \textit{Gaia} will
also provide a much better picture of the completeness of the samples.  For
the SDSS sample in particular, this includes the identification of double
degenerates and the determination of precise masses for all subtypes including
magnetic white dwarfs. By identifying a much larger 40~pc sample with the help
of \textit{Gaia} and spectroscopic follow-ups, it will be possible to improve
our understanding of the local SFH and kinematics as a function of age and
mass.  For instance, it will be possible to study the mass distribution for
subsamples in total age, reducing the uncertainties due to Galactic evolution
effects. Nevertheless, it could remain a challenge to disentangle the effects
from the IMF and IFMR on the mass distribution even with the \textit{Gaia}
data, although many more white dwarfs in clusters and common proper motion
pairs will be discovered allowing to improve the IFMR.  Our study will be
useful to re-assess all uncertainties in the \textit{Gaia} era.

\subsection{Constraining the Merger Population}

Little is known about the fraction of white dwarfs that are the product of
mergers (WD+WD, WD+RG, or RG+RG, where RG stands for red giant) in the solar
neighborhood. \citet{wegg12} have analysed the kinematics of massive SDSS
white dwarfs and demonstrated that they have the characteristics of a young
singly-evolved population.  From multi-epoch spectroscopy of SDSS white
dwarfs, \citet{badenes12} have calculated the WD+WD merger rate to be around
$1.4 \times 10^{-13}$ yr$^{-1} M_{\odot}^{-1}$.  Binary population synthesis
models predict merger rates that are about twice as large
\citep{iben97,toonen12}. For the last 10 Gyr, this leads to approximately one
merger product in our main SDSS subsample in the range $16,000 < T_{\rm eff}$
(K) $<$ 22,000.  On the other hand, this does not include stars that have
merged before both of them became white dwarfs (WD+RG or RG+RG), which could
account for a larger fraction. However, there is little evidence that a
merging process involving red giants would favour the production of white
dwarfs that are more massive than the average \citep{berro12}.

On the other hand, early investigations of the white dwarf mass distribution
have identified a high-mass peak or so-called bump around 0.8 $M_{\odot}$
\citep{marsh97,vennes99}, and proposed a merger population as a possible
cause. More recently, \citet{gia12} and \citet{kleinman13} have also suggested
that the high-mass peak is likely due to mergers. However, these studies are
not based on extensive simulations of stellar populations, which for instance
\citet{kleinman13} are cautious to mention. Nevertheless, \citet{gia12}
suggests that mergers account for $\sim$3\% of the 20~pc sample. This is much
larger than the observed WD+WD merger rate.

From the simulations performed in this work, we can suggest a number of
alternatives to explain features in the field white dwarf mass distribution.
First of all, our standard set of simulations already predict too many massive
white dwarfs. Hence there is no need to invoke mergers to explain even the
most massive (non-magnetic) white dwarfs in the current samples. This applies
to the SDSS mass distribution at all temperatures according to
Table~\ref{fg:t2}. Furthermore, even our standard simulations have a bump
around 0.8 $M_{\odot}$, and this is most easily seen when we neglect
observational errors in Fig.~\ref{fg:f10}. This feature is in fact due to the
two-piece polynomial fit to the IFMR of \citet{cummings16}. It is not present
when using any linear IFMR. We do not claim that it is a real astrophysical
feature of the IFMR even though our parameterisation was motivated by
theoretical IMFRs \citep{cummings16}.  It merely demonstrates that current
constraints on single star evolution and the IFMR do not rule out the presence
of a high-mass peak.  Finally, there are number of biases that impact the
field white dwarf mass distribution, the combination of which could cause the
high-mass peak. We conclude that no evidence of WD+WD mergers can be found in
the field white dwarf mass distribution.

\section{Conclusions}

We have presented a thorough study on the astrophysical interpretation of the
field white dwarf mass distribution. We have chosen the well studied 20~pc and
SDSS samples, restricting the latter to $16,000 < T_{\rm eff}$ (K) $<$ 22,000,
$16 < g < 18.5$, and single non-magnetic white dwarfs in order to have a well
understood completeness. Our approach has been to perform Monte Carlo
simulations to compare with the observations. The first result of this work is
that we predict a larger mean mass for the 20~pc sample in comparison to the
SDSS sample, in agreement with the observations. This suggests that the
photometric technique and model atmospheres of cool white dwarfs, largely
employed for the local sample, are in agreement with the results of the
spectroscopic technique for hotter DA and DB white dwarfs is the SDSS.

Our simulations reproduce reasonably well the samples studied in this work
using standard assumptions about stellar and Galactic evolution. However, for
both samples our simulations predict too many high-mass white dwarfs ($M >
0.75$ $M_{\odot}$) by a factor of $\sim$1.5. From our extensive review of
biases that impact our simulations, we find that this offset is not
unexpected. Probable causes are uncertainties in the assumed IFMR, IMF, SFH,
variation of Galactic disk vertical scale height as a function of total
stellar age, binary evolution, neglect of magnetic white dwarfs (SDSS), and
unidentified faint massive objects (20~pc sample). While a majority of these
uncertainties will be improved with \textit{Gaia}, it could remain a challenge
to disentangle the effects from the IFMR and IMF.

Our results challenge the interpretation that there is evidence for a
population of WD+WD mergers in the field white dwarf mass distribution. On the
contrary, we find no observed excess of high-mass objects and features in the
observed distributions can not be unambiguously linked to mergers. We note
that our results do not rule out a population WD+MS or WD+WD mergers that are
not preferentially massive, or that some percentage of known massive single
white dwarfs, e.g. with large magnetic fields, could be directly linked to a
merger event.

\section*{Acknowledgements}

Support for this work was provided by NASA through Hubble Fellowship grant
\#HF-51329.01 awarded by the Space Telescope Science Institute, which is
operated by the Association of Universities for Research in Astronomy, Inc.,
for NASA, under contract NAS 5-26555. This project was supported by the
National Science Foundation (NSF) through grant AST-1211719. We would like
to thank Jay Holberg for commenting on our manuscript.

\bsp	% typesetting comment
\label{lastpage}
\end{document}